\documentclass[preprint,floats,tightenlines,11pt,eqsecnum,showpacs,aps]{revtex4}
\usepackage{amsmath}
\usepackage{graphicx}
\begin{document}
\def\appls{\hbox{$<$\kern-.75em\lower 1.00ex\hbox{$\sim$}}}

%\title{STUDY OF MEASUREMENTS OF $\pi^- p \to \pi^- \pi^+ n$ ON POLARIZED TARGET\\ I. EVIDENCE FOR VIOLATION OF UNITARY EVOLUTION LAW\\IN HADRON SCATTERING}

%\title{CONSTRAINTS ON AMPLITUDES IN $\pi N \to \pi \pi N$ SCATTERING IMPOSED BY UNITARY EVOLUTION LAW AND THEIR VIOLATION BY CERN MEASUREMENTS OF $\pi^- p \to \pi^- \pi^+ n$ ON POLARIZED TARGET}

%\title{EVIDENCE FOR $\rho^0(770)-f_0(980)$ MIXING AND THE VIOLATION OF UNITARY EVOLUTION LAW IN MEASUREMENTS OF $\pi^- p \to \pi^- \pi^+ n$ ON POLARIZED TARGET} 

%\title{EVIDENCE FOR THE VIOLATION OF UNITARY EVOLUTION LAW IN MEASUREMENTS OF $\pi^- p \to \pi^- \pi^+ n$ ON POLARIZED TARGET} 

%\title{UNITARY AND NON-UNITARY EVOLUTION LAWS IN $\pi N \to \pi \pi N$ SCATTERING ON POLARIZED TARGET} 

%\title{TEST OF THE UNITARY EVOLUTION LAW IN CERN MEASUREMENTS OF $\pi^- p \to \pi^- \pi^+ n$ ON POLARIZED TARGET AND THE EVIDENCE FOR A QUANTUM ENVIRONMENT AND ITS INTERACTION WITH PARTICLE SCATTERING} 

\title{STUDY OF $\pi N \to \pi \pi N$ PROCESSES ON POLARIZED TARGETS:\\
QUANTUM ENVIRONMENT AND ITS DEPHASING INTERACTION WITH PARTICLE SCATTERING}

\author{Miloslav Svec\footnote{electronic address: svec@hep.physics.mcgill.ca}}
\affiliation{Physics Department, Dawson College, Montreal, Quebec, Canada H3Z 1A4}
%\date{April 23, 2013}
%\date{July 13, 2014}
%\date{November 27, 2014}
\date{March 24, 2015}

\begin{abstract}

Unitary evolution law describes isolated particle scattering processes in an empty Minkowski spacetime. We put forward a hypothesis that the physical Universe includes a quantum environment that interacts with some particle scattering and decay processes. While the scattering process is governed by the $S$-matrix dynamics and its conservation laws and unitarity, the interaction with the environment evolves the produced final state $\rho_f(S)$ to the observed state $\rho_f(O)$. To be consistent with the Standard Model this new interaction must be a pure dephasing interaction. Governed by a non-unitary evolution law, it modifies the phases of the $S$-matrix amplitudes and can give rise to mixing of such amplitudes to form observed amplitudes.

We present the first test of unitary evolution law in particle scattering.
Conservation of $P$-parity in strong interactions imposes constraints on partial wave helicity and nucleon transversity amplitudes in $\pi N\to \pi \pi N$ processes. An independent set of constraints on these amplitudes is imposed by the $S$-matrix unitary evolution law. The unitary evolution evolves pure initial states into pure final states leading to 9 independent constraints on 16 components of angular intensities in $\pi N \to \pi \pi N$ processes. When expressed in terms of parity conserving transversity amplitudes, all 9 constraints are identities provided a single constraint on the transversity amplitudes holds true. The constraint implies that relative phases between transversity amplitudes of the same naturality and transversity must be $0$ or $\pm \pi$. Assuming a self-consistent set of these unitary phases we use the CERN data on spin observables $R^0_u$ and $R^0_y$ to determine a unique solution for the $S$- and $P$-wave moduli below 1080 MeV. The data require $\rho^0(770)-f_0(980)$ mixing in the $S$-wave but this unitary solution is excluded by data on observables $R^0_x$ within at least 5 standard deviations. All previous amplitude analyses of $\pi N \to \pi \pi N$ processes found non-unitary relative phases in an apparent violation of the unitary evolution law. The contrast between the predicted unitary relative phases and the observed non-unitary phases presents an unambigous evidence for the non-unitary evolution of the produced final state and supports the hypothesis of the existence of a quantum environment and its pure dephasing interaction with particle scattering processes.

\end{abstract}

\maketitle

\tableofcontents

\newpage
\section{Introduction.}

The concept of the $S$-matrix is deeply rooted in the concept of Minkowski spacetime and its Poincare symmetry. The Poincare group on the Minkowski spacetime allows us to define particle four-momentum and spin. This in turn allows us to consider particle scattering and decay of incident states into outgoing states. The resulting probability amplitudes are summarized as $S$-matrix elements. The conservation of probability imposes unitarity of the $S$-matrix. The $S$-matrix commutes with the generators of the Poincare group. As a result of this symmetry the total four-momentum and total angular momentum are conserved in particle scattering and decays. Internal symmetries of the $S$-matrix impose additional conservation laws. 

In $S$-matrix theory particle scattering and decays are isolated and 
time-reversible quantum events in Minkowski spacetime. The reason for this is that Minkowski spacetime is empty. There is no quantum environment in Minkowski spacetime with which the scattering and decay processes could interact. Another reason is that the scattering and decay processes do not interact with the Minkowski spacetime itself since it has no quantum structure that, in effect, could present itself as a quantum environment.

Particle scattering and decay processes take place, in fact, in real physical Universe. Suppose that there exists a quantum environment in the Universe that interacts with particle processes. Such interaction cannot originate in the known interactions of the Standard Model. These interactions would lead to observable violation of the conservation laws and render the dynamics of particle interaction inaccessible to experiment. The interaction of particle processes with the quantum environment must originate from the outside of the Standard Model. If the quantum environment and this new kind of interaction are to be an integral part of the Nature, then they must be fully consistent with the conservation laws and unitarity of the Standard Model. 

There exists such an interaction in the Nature. It is the pure dephasing interaction between a quantum system $S$ and a quantum environment $E$ which is a non-dissipative interaction that affects only the phase(s) of the quantum system $S$. In general, its effect is to change the quantum information content of the quantumm system $S$. Unlike the force of gravity, which has undetectable effects on particle processes, the pure dephasing interaction could be observable. The key to observing dephasing interaction, and thus the quantum environment, is the $S$-matrix unitarity.

The evolution of an isolated initial state $\rho_i$ into isolated final states $\rho_f(S)$ is governed by the unitary evolution law 
\begin{equation}
\rho_f(S)=S\rho_iS^+ 
\end{equation}
Evolution of an initial state $\rho_i$ interacting with a quantum environment $E$ is governed by a non-unitary evoluion law given by Kraus representation
\begin{equation}
\rho_f(A) = \sum \limits_k A_k \rho_i A_k^+
\end{equation}
where $A_k$ are unitary or non-unitary Kraus operators~\cite{kraus71,kraus83,nielsen00,bengtsson06}. The necessary and sufficient condition for the non-unitary evolution law to be consistent with the unitary evolution law is that the interaction with the environment be a final state interaction which involves only the produced state $\rho_f(S)$. The non-unitary evolution of the $S$-matrix final state $\rho_f(S)$ yields the observed non-unitary final state $\rho_f(O)$
\begin{equation}
\rho_f(O) = \sum \limits_k A_k \rho_f(S) A_k^+
\end{equation}
The effect of the pure dephasing interaction with the environment thus shall be to modify the phases of the $S$-matrix amplitudes in the observed amplitudes of the new state $\rho_f(O)$. The existence of the quantum environment will manifests itself in the difference between the phases of the observed and $S$-matrix amplitudes. To an observer (initially) unaware of the non-unitary evolution of the produced final state $\rho_f(S)$ this difference would appear as an apparent violation of the unitary evolution law. An observer who insists on the validity of the unitary evolution law would explain the difference differently: there is a presence of a short time scale in which Standard Model interactions act to produce the $S$-matrix final state $\rho_f(S)$ and afterwards a non-unitary evolution of this state leaves its imprints on the observed final state $\rho_f(O)$.

To gain information about the phases of the $S$-matrix amplitudes we shall use the fact that the $S$-matrix unitary evolution law evolves pure initial states $\rho_i$ into pure final states $\rho_f(S)$ in exclusive processes. In two-body processes such as $\pi N \to \pi N$ this condition imposes no constraints on the amplitudes and thus no specific information about their phases. The situation is different in $\pi N \to \pi \pi N$ and similar production processes.

In this work we aim to test the unitary evolution law in the pion production $\pi^- p \to \pi^- \pi^+n$ measured at CERN on polarized target at 17.2 GeV/c~\cite{rybicki96}. We develop the necessary spin formalism and show that the purity of the final state density matrix in $\pi N \to \pi \pi N$ processes is controlled by the recoil nucleon polarization. Evolution of pure initial states to pure final states imposes 9 constraints on 16 angular intensities describing the final state. Using $P$-parity conservation we show that all constraints are identities provided that the partial wave amplitudes satisfy the conditions
\begin{equation}
Im(U^J_{\lambda \tau}N^{K*}_{\mu -\tau})=0
\end{equation}
where $U^J_{\lambda \tau}$ and $N^K_{\mu -\tau}$ are parity conserving nucleon transversity unnatural and natural exchange amplitutes with dipion spin, helicity and nucleon transversity $J,\lambda,\tau$ and $K,\mu,-\tau$, respectively. The conditions (1.4) imply that the relative phases between any two unnatural or two natural exchange amplitudes of the same transversity $\tau$ as well as between any unnatural and natural exchange amplitudes of opposite transversities must be $0$ or $\pm \pi$.  As a result, the unnatural and natural amplitudes share a common phase $\Phi(U^0_{0\tau})$ and $\Phi(N^1_{1\tau})$, respectively. Because strong, electromagnetic and weak interactions do not mix particles with different spins, self-consistency requires that there be no mixing of resonances of different spins in any partial wave amplitudes with such unitary phases. 

The elegant simplicity and uniqueness of these predictions render the test of unitary evolution law possible using the existing CERN measurements on polarized target at dipion masses below 1080 MeV where $S$- and $P$-wave amplitudes dominate. We found that the measurements of density matrices $R^0_u$ and $R^0_y$ yield a solution with $\rho^0(770)-f_0(980)$ mixing in the $S$-wave. But this solution is entirely excluded by the measured data on density matrix $R^0_x$ within at least five standard deviations. This result shows that the data require $\rho^0(770)-f_0(980)$ mixing in the $S$-wave but reject unitary phases in an apparent violation of the unitary evolution law.

All previous amplitude analyses of $\pi^- p \to \pi^- \pi^+  n$ at 17.2 GeV/c  ~\cite{becker79a,becker79b,chabaud83,rybicki85,svec92a,svec96,svec97a,svec07b,svec07c,svec12a} and at 1.78 GeV/c~\cite{alekseev99} as well as of $\pi^+ n \to \pi^+ \pi^- p$ at 5.98 and 11.85 GeV/c~\cite{svec92a,svec96,svec97a} found non-unitary relative phases of all transversity amplitudes. Recent amplitude analysis of $S$- and $P$-wave subsystem in $\pi^- p \to \pi^- \pi^+  n$ at 17.2 GeV/c established that the width of $\rho^0(770)$ resonance peak observed in all $P$-wave amplitudes does not depend on its helicity $\lambda$~\cite{svec07b,svec12a} as required by the rotational/Lorentz symmetry of the $S$-matrix. Furthermore, the non-unitary relative phases of the $S$-and $P$-wave amplitudes are near the unitary values. These findings show that the observed non-unitary phases are consistent with the $S$-matrix unitary evolution law for the production process. This is possible if the non-unitary phases arise from a pure dephasing interaction of the produced $S$-matrix final state $\rho_f(S)$ with a quantum environment with. The contrast between the predicted unitary relative phases and the observed non-unitary phases therefore presents an unambigous evidence for the non-unitary evolution of the produced final state and supports the hypothesis of the existence of the quantum environment and its pure dephasing interaction with particle scattering processes.

In a sequel paper~\cite{svec13b} we show that the consistency of the pure dephasing interaction with the Standard Model in $\pi N \to \pi \pi N$ processes requires that it be a dipion spin mixing interaction the effect of which is the mixing of $S$-matrix partial wave amplitudes to form observable amplitudes. The theory predicts  $\rho^0(770)-f_0(980)$ mixing in the $S$-and $P$-wave amplitudes in $\pi^- p \to \pi^- \pi^+ n$. The predicted moduli and relative phases of the mixed amplitudes are in an excellent qualitative agreement with the experimental results~\cite{svec07b,svec12a}. The evidence for $\rho^0(770)-f_0(980)$ mixing dates back to 1960's~\cite{hagopian63,islam64,patil64,durand65,baton65,donohue79} and was later confirmed in all amplitude analyses  of $\pi^- p \to \pi^- \pi^+  n$ and $\pi^+ n \to \pi^+ \pi^- p$ on polarized targets. A survey of evidence for $\rho^0(770)-f_0(980)$ mixing from all these amplitude analyses on polarized targets is presented in Ref.~\cite{svec12d}.

The issue of the experimental test of the unitary evolution law was first raised in 1974 by Marinov who suggested to describe the time evolution of $K^0 \overline{K^0}$ system using a model of a non-unitary evolution from  pure states into mixed states~\cite{marinov74}. He found that such evolution is time-irreversible and violates $CPT$ symmetry and proposed to make a complete measurement of matrix elements of $\rho_f$ to test the unitary evolution law. In 1980 Wald showed rigorously that any scattering process of particles that evolves pure initial state into a mixed final state is not an invertible process and therefore it is time-irreversible and violates $CPT$ symmetry~\cite{wald80}. He suggested that such non-unitary processes will occur in curved space-time, and that quantum gravity violates $CPT$ symmetry and time-reversal invariance. In 1982 Hawking pointed out that particle scattering does not take place in a structureless continuum of the Minkowski space-time but in an environment of quantum space-time fluctuations and suggested that pure initial states of interacting particles will evolve into mixed final states due to the interaction of the particle scattering process with quantum fluctuations of the space-time metric  - at any energy~\cite{hawking82,hawking84}. Hawking questioned the universal validity of the unitary time evolution in the presence of metric fluctuations, and suggested that initial and final state density matrices $\rho_{i}$ and $\rho_{f}$ are connected by a non-unitary evolution law described by a linear but non-unitary and non-invertible superscattering operator
\begin{equation}
\rho_f= \text{\it{\$}} \rho_i
\end{equation}
To avoid negative probabilities the mapping (1.5) must be completely positive. A linear mapping (1.5) is completely positive if and only if it has the form of the Kraus representation (1.2)~\cite{kraus71,kraus83,nielsen00,bengtsson06}. Recently Unruh and Wald~\cite{unruh95} and Oppenheim and Reznik~\cite{oppenheim09} demonstrated the feasibility of non-unitary evolution in quantum field theories. However in these models Lorentz invariance fails. In 2002 Greenberg showed that any local interacting (scattering producing) theory that violates $CPT$ invariance necessarily violates Lorentz invariance~\cite{greenberg02}. 

Hawking's ideas inspired attempts to test unitary evolution law experimentally. The efforts focused mainly on non-unitary time evolution of neutral kaons $K^0 \overline{K^0}$ system using Lindblad type evolution laws~\cite{ellis84,ellis96,huet95,benatti97,gerber98,gerber04} and led to the predictions of $CPT$ violation and a modification of EPR (Einstein-Podolsky-Rosen) correlations~\cite{bernabeu06a,bernabeu06b}. During the recent years experiments with neutral kaons have yielded sensitive results on violations of $CPT$ symmetry, time reversal invariance and entanglement of kaon pairs~\cite{fidecaro06,ambrosino06,amelino10}. So far these experiments did not provide a conclusive confirmation of a non-unitary evolution of free neutral kaon systems. These observations can be understood as the result of the absence of the diparticle spin mixing in the time evolution of the $K^0 \overline{K^0}$ system.

What are the implications of the non-unitary evolution for the $CPT$ symmetry in $\pi N \to \pi \pi N$ processes? In Quantum Field Theory and in the Standard Model $CPT$ symmetry is conserved as a consequence of locality, Lorentz symmetry and Hermitian interaction Lagrangian~\cite{streater80,sachs87}. This means that the produced final state $\rho_f(S)$ and the $S$-matrix amplitudes are $CPT$ symmetric. According to Wald Theorem non-unitary evolution of $\rho_f(S)$ to the observed state $\rho_f(O)$ violates $CPT$ symmetry and is time irreversible. Greenberg Theorem then implies a violation of Lorentz symmetry. This means that the non-unitary evolution and pure dephasing interaction cannot be described by local and Lorentz invariant quantum field theory. In Ref.~\cite{svec13b} we show that the Kraus operators are forward scattering operators akin to forward scattering of light in a refracting medium. The forward scattering of dipion spin states on recoil nucleon into dipion spin states with different spin leads to dephasing and spin mixing of the $S$-matrix amplitudes to form the observed amplitudes.

Interactions that are invariant under $CPT$ symmetry lead to observables that are invariant under $CPT$ as well. Conversely, interactions that violate $CPT$ symmetry lead to observables that violate $CPT$ symmetry. It is important to recognize that the $CPT$ violating non-unitary evolution and interaction with the quantum environment do not contradict the $CPT$ invariant interactions and their observables involved in the production of the state $\rho_f(S)$. Thus we may expect e.g. the masses and lifetimes of $\pi^-$ and $\pi^+$ to be the same and rotationally invariant witdth and mass of the $\rho(770)$ resonance in $\pi^- p \to \pi^-\pi^+n$. The $CPT$ violating interactions will manifest themselves in other observable aspects of the pion creation process. In this case the $CPT$ violating observables are the observed transversity amplitudes. The distinct $CPT$ violating effects seen only in these non-unitary amplitudes are their non-unitary phases and the spin mixing. As an alternative to spin mixing certain observed amplitudes violate so called cosine gap condition which $S$-matrix amplitudes must satisfy~\cite{svec13b}.

The paper is organized as follows. In Section II. we briefly discuss the evolution of pure initial states by unitary and non-unitary evolution laws. In Section III. we show that $S$-matrix unitary evolution implies evolution from pure initial states into pure final states in exclusive processes. In Sections IV.-VI. we review and develop the spin formalism necessary to derive and discuss the constraints imposed by the unitary evolution law on the nucleon transversity amplitudes. We emphasize the coherent superposition of diparticle states and the constraints imposed by the conservation of $P$-parity on partial wave amplitudes. Final state density matrix is defined and its properties derived in some detail. This material will be used also in the sequel paper. In Section VII. we present the unitary constraints on angular intensities from which we derive the unitary condition (1.4) and a self-consistent set of unitary phases. In Section VIII. we present a unique unitary solution for $S$- and $P$-amplitudes below 1080 MeV and show it is excluded by the data on observables $R^0_x$. This conclusion does not change when we include $D$-waves in the analysis. We discuss the evidence for the quantum environment and the dephasing interaction in Section IX. and a physical interpretation of the unitary and non-unitary relative phases in Section X.. We present our conclusions and outlook in the Section XI.. The Appendix provides an outline of the proof of the unitary condition (1.4).

%\newpage
\section{Unitary and non-unitary evolution laws.}

Let $H$ be a Hilbert space of vector states with an orthonormal basis $|n>, n=1,N$ where $N=\dim H$. Let $\mathcal{B}(H)$ be the Hilbert-Schmidt space of linear operators on $H$ with an orthonormal basis $|n><m|, n,m=1,N$. Density matrix $\rho \in \mathcal{B}(H)$  is a hermitian and positive operator $\rho = \sum \rho_{m n}|m><n|$ with $Tr(\rho)=1$. Density matrix $\rho$ represents a pure state if it satisfies condition $Tr(\rho^2)=(Tr(\rho))^2$.
A quantum state satisfies this purity condition if and only if its density matrix has the form $\rho=|\Psi><\Psi|$ where $|\Psi>$ is a vector in $H$.

Let $A$ be a linear operator on $H$ and $\rho_i$ an arbitrary density matrix in $\mathcal{B}(H)$. Then $A$ defines a mapping - or an evolution law - of the state $\rho_i$ into a state $\rho_f$
\begin{equation}
\rho_f=A \rho_i A^+
\end{equation}
Let $\rho_i=|i><i|$ be a pure state and $A|i>=|\Psi_A(i)>$. Then $\rho_f=A|i><i|A^+=|\Psi_A(i)><\Psi_A(i)|$ is also a pure state. If $A$ is a unitary operator the mapping is a trace conserving and invertible unitary evolution. Otherwise it is a simple non-unitary evolution with normalized density matrix $\rho_f'=\rho_f/Tr(\rho_f)$.

To be physically meaningful the mappings from initial to final states must preserve the positivity of all probabilities which requires that they be completely positive. The necessary and sufficient condition for an evolution law to be completely positive is that it has the form called Kraus representation~\cite{kraus71,kraus83,nielsen00,bengtsson06}
\begin{equation}
\rho_f = \sum \limits_k A_k \rho_i A_k^+
\end{equation}
where $A_k$ are unitary or non-unitary Kraus operators. For trace preserving evolution they satisfy completness relation
\begin{equation}
\sum \limits_k A_k^+ A_k = I
\end{equation}
Kraus representation (2.2) describes a general completely positive, non-unitary and non-invertible evolution. Let $\rho_i=|i><i|$ be a pure initial state. Then the final state is a mixed state 
\begin{equation}
\rho_f=\sum \limits_k A_k|i><i|A_k^+=\sum \limits_k |\Psi_{A_k}(i)><\Psi_{A_k}(i)|
\end{equation}
A special case of (2.2) is the evolution law (2.1) which evolves pure initial states into pure final states. It describes evolution of isolated quantum systems while the Kraus representation describes a non-unitary evolution of open quantum systems. Kraus representation arises from a unitary evolution law governing the co-evolution of the quantum system with its quantum environment after the interacting degrees of freedom between the two systems have been traced out in their joint final state density
matrix~\cite{kraus71,kraus83,nielsen00,bengtsson06}. The dissipative dephasing interactions exchange not only phases but also four-momentum and/or angular momentum. There is no exchange of four-momentum and/or angular momentum in pure dephasing interactions of the quantum system with its quantum environment. 

\section{Evolution from pure initial states into pure final states in\\ exclusive processes.}

Unitary $S$-matrix evolves an arbitrary initial state $\rho_i$ into a final state
\begin{equation}
\rho=S\rho_iS^+
\end{equation}
Any pure initial state $\rho_i$ can be written in the form $\rho_i=|i><i|$. The evolution operator $S$ brings the state vector $|i>$ to a state vector $|\Psi>=S|i>$ so that $\rho=|\Psi><\Psi|$ is a pure state. Using a completness relation
\begin{equation}
\sum \limits_f \sum \limits_{\chi_f} \int d\Phi_f |p_f,\chi_f,\gamma_f><p_f,\chi_f,\gamma_f|=I
\end{equation}
the state vector $|\Psi>$ has an explicit form
\begin{equation}
|\Psi>=\sum \limits_f \sum \limits_{\chi_f} \int d\Phi_f |p_f,\chi_f,\gamma_f><p_f,\chi_f,\gamma_f|S|i>
\end{equation}
where the first sum is over all allowed final states $f$, the second sum is over final state spins $\chi_f$ and the integration is over the entire phase space of final state momenta $p_f$. The symbol $\gamma_f$ labels the quantum numbers of the state $f$. The density matrix $\rho=|\Psi><\Psi|$ has an explicit form 
\begin{equation}
\rho=\sum \limits_f \rho_f +\sum \limits_{f'} \sum \limits_{f''\neq f'} \rho_{f'f''}
\end{equation}
where
\begin{eqnarray}
\rho_f & = & \sum \limits_{\chi_f',\chi_f''} \int d\Phi_f' d\Phi_f'' 
<p_f',\chi_f',\gamma_f|S|i><i|S^+|p_f'',\chi_f'',\gamma_f>
|p_f',\chi_f',\gamma_f><p_f'',\chi_f'',\gamma_f|\\
\nonumber
\rho_{f'f''} & = & \sum \limits_{\chi_{f'},\chi_{f''}} \int d\Phi_{f'} d\Phi_{f''} 
<p_{f'},\chi_{f'},\gamma_{f'}|S|i><i|S^+|p_{f''},\chi_{f''},\gamma_{f''}>
|p_{f'},\chi_{f'},\gamma_{f'}><p_{f''},\chi_{f''},\gamma_{f''}|
\end{eqnarray}
The density matrix $\rho_f$ of the final state $f$ is the block-diagonal submatrix of $\rho$. It is a pure state $\rho_f=|\Psi_f><\Psi_f|$ where
\begin{equation}
|\Psi_f>=\sum \limits_{\chi_f} \int d\Phi_f |p_f,\chi_f,\gamma_f><p_f,\chi_f,\gamma_f|S|i>
\end{equation}
The projection of $\rho_f$ into a state $\rho_f(p_f)$ with definite final state momenta $p_f$ is given by
\begin{equation}
\rho_f(p_f)=|p_f,\gamma_f><p_f,\gamma_f|\rho|p_f,\gamma_f><p_f,\gamma_f|=
\end{equation}
\[
\sum \limits_{\chi_f',\chi_f''}<p_f,\chi_f',\gamma_f|S|i><i|S^+|p_f,\chi_f'',\gamma_f>
|p_f,\chi_f',\gamma_f><p_f,\chi_f'',\gamma_f|
\]
It is a pure state $\rho_f(p_f)=|\Psi_f(p_f)><\Psi_f(p_f)|$ where
\begin{equation}
|\Psi_f(p_f)>=\sum \limits_{\chi_f} |p_f,\chi_f,\gamma_f><p_f,\chi_f,\gamma_f|S|i>
\end{equation}
Note that $\sum \limits_f \int d\Phi_f |p_f,\gamma_f><p_f,\gamma_f|=I$ since the spin projection operators $\sum \limits_{\chi_f} |\chi_f><\chi_f|=I$. The $S$-matrix unitary evolution (3.1) and the completness relation (3.2) thus imply that for any initial pure state $\rho_i$ all final states $\rho_f(p_f)$ must be pure states. 

\section{Two-particle coherent states and diparticles.}

The state vector $|p_1 p_2; \mu_1 \mu_2; \gamma>$ of two non-interacting particles with four-momenta $p_1,p_2$, helicities $\mu_1, \mu_2$ and quantum numbers $\gamma$ is a direct product of two single-particle helicity states. It is an eigenstate of the momentum operator $P_\mu$ with eigenvalue $p=p_1+p_2$ and invariant mass $m^2=p^2$. It does not define an irreducible representation of the Restricted Inhomogenous Lorentz group, and therefore it has no definite spin and $P$-parity. However, it can be expressed as a coherent superposition of spin states with four-momentum $p$ and definite spin and parity. These states have a character of free non-interacting single particle helicity states with variable mass and carry the quantum numbers $\gamma$. We shall refer to these states as diparticles.

In the center-of-mass system where $\vec{p^*}=\vec{p_1^*}+\vec{p_2^*}=0$ and $E^*=m$ the two-particle coherent state reads~\cite{jacob59,martin70}
\begin{equation}
|p_1^*p_2^*;\mu_1\mu_2;\gamma>=\bigl ({4m\over{q}}\bigr )^{1\over{2}}|p^*>|\theta \phi;\mu_1\mu_2;\gamma>=
\end{equation}
\[
\bigl ({4m\over{q}}\bigr )^{1\over{2}}\sum \limits_{J \lambda} \sqrt{{2J+1} \over {4\pi}} D^J_{\lambda,\mu}(\phi,\theta,-\phi)|p^*> |J\lambda;\mu_1 \mu_2;\gamma>
\]
where $\mu=\mu_1-\mu_2$ and $q=q(m^2)$, $\theta,\phi$ describe the momentum and the direction of the particle 1 in the center-of-mass system. The summation is over all integral or half-integral values of $J$. In the two-particle rest frame $\vec{p^*}=0$ and $\lambda$ is the component of the spin $J$ along the direction of the $z$-axis. As in the case of single particle spin states, a boost along the $z$-axis and a rotation define a pure Lorentz transformation $\Lambda_p$ that brings the state $|p^*> |J\lambda;\mu_1 \mu_2;\gamma>$ from the rest frame to a state with any momentum $p$ on the orbit
\begin{equation}
U(\Lambda_p)\bigl (|p^*>|J\lambda;\mu_1 \mu_2;\gamma>\bigr )=
|p>|J\lambda;\mu_1 \mu_2;\gamma>
\end{equation} 
Here $\lambda$ is now a helicity of the angular helicity state $|p>|J\lambda;\mu_1 \mu_2;\gamma>$ in the direction of $\vec{p}$. The states $|p>|J\lambda;\mu_1 \mu_2;\gamma>$ define irreducibble representation of the Restricted Inhomogehous Lorentz group. Under any element of the group they transform accordingly
\begin{equation}
U(\Lambda,a)\bigl (|p>|J\lambda;\mu_1 \mu_2;\gamma> \bigr)=
\end{equation}
\[
\exp{(-ip_\mu^{'}a^\mu)}|p'>\Bigl (\sum \limits_{\lambda'} D^J_{\lambda',\lambda}(R) |J\lambda';\mu_1 \mu_2;\gamma> \Bigr )
\]
where $p'=\Lambda p$, $R=\Lambda^{-1}_{p'}\Lambda \Lambda_p$ is a Wigner rotation~\cite{martin70,perl74} and $D^J_{\lambda',\lambda}(R)$ is the matrix representing the rotation $R$ in the irreducible representation of the rotation group corresponding to spin $J$.

Intrinsic $P$-parity of single particle states is defined in their rest frame. Similarly, the intrinsic $P$-parity of the angular helicity states is given by a relation in the center-of-mass system~\cite{martin70,jacob59}
\begin{equation}
P(|p^*>|J\lambda;\mu_1 \mu_2;\gamma>)=\eta_1 \eta_2 (-1)^{J-s_1-s_2}|p^*>|J\lambda;-\mu_1-\mu_2;\gamma>
\end{equation}
where $\eta_1,\eta_2$ and $s_1,s_2$ are the parities and spins of the two particles, respectively. The angular helicity states are parity eigenstates only for two spinless particle states. However we can write any angular helicity state as a combination of two states with opposite $P$-parities
\begin{equation}
|J\lambda;\mu_1 \mu_2;\gamma>={1\over{2}}\bigl (|J\lambda+;\mu_1 \mu_2;\gamma>+ |J\lambda-;\mu_1 \mu_2;\gamma> \bigr )
\end{equation}
where
\begin{equation}
|J\lambda\pm;\mu_1 \mu_2;\gamma>=
|J\lambda;\mu_1 \mu_2;\gamma> \pm 
\eta_1 \eta_2 (-1)^{J-s_1-s_2}|J\lambda;-\mu_1-\mu_2;\gamma>
\end{equation}
The angular helicity states with a definite $P$-parity now have a character of single-particle helicity states with variable mass $m$ and quantum numbers $\gamma$. We can refer to these states as diparticle spin states. The general two-particle helicity states are a coherent superposition of diparticle spin states
\begin{equation}
|p_1p_2;\mu_1\mu_2;\gamma>=\bigl ({4m\over{q}}\bigr )^{1\over{2}}|p>|\theta \phi;\mu_1\mu_2;\gamma>=
\end{equation}
\[
\bigl ({4m\over{q}}\bigr )^{1\over{2}} \sum \limits_{J \lambda} \sqrt{{2J+1} \over {4\pi}} D^J_{\lambda,\mu}(\phi,\theta,-\phi){1\over{2}}\Bigl (|p>|J\lambda+;\mu_1 \mu_2;\gamma> +|p>|J\lambda-;\mu_1 \mu_2;\gamma> \Bigr )
\]
For two particles with spin the coherent state $\bigl ({4m\over{q}}\bigr )^{1\over{2} }|p>|\theta \phi;\mu_1\mu_2;\gamma>$ is an angular superposition of diparticle states that form parity doublets. For spinless particles the diparticle states are parity singlets. An extension of this spin formalism for three-particle helicity states was given by Wick~\cite{wick62}.

\section{Amplitudes in $\pi N \to \pi \pi N$ processes.}

\subsection{Partial wave helicity amplitudes}

We consider pion creation process $\pi_a N_b \to \pi_1 \pi_2 N_d$ with four-momenta $p_a+p_b = p_1+p_2+p_d$. In the laboratory system of the reaction the $+z$ axis has the direction opposite to the incident pion beam. The $+y$ axis is perpendicular to the scattering plane and has direction of $\vec {p}_a \times \vec {p}_c$ where $p_c=p_1+p_2$. The angular distribution of the produced dipion system is described by the direction of $\pi_1$ in the two-pion center-of-mass system and its solid angle $\Omega = \theta, \phi$. The final state vector for the non-interacting particles is
\begin{equation}
|p_10>|p_20>|p_d \chi>=\bigl ({4m \over{q}}\bigr )^{1\over{2}}|p_c>|\theta \phi,00>|p_d\chi>
\equiv \bigl ({4m \over{q}}\bigr )^{1\over{2}}|p_c p_d>|\theta \phi,\chi>
\end{equation}
where $\chi$ is the recoil nucleon helicity, $m$ is the invariant mass $m^2=p_c^2$ and $q=q(m^2)$ is the $\pi_1$ momentum in the two-pion center-of-mass system. The helicity of target nucleon is $\nu$. We have seen in Section IV. that a state vector of two non-interacting particles can be expressed as a coherent superposition of diparticle helicity states with definite spin and parity given by (4.7). For two pions $\mu_1=\mu_2=0$ and $D^J_{\lambda0}(\phi,\theta,-\phi)=\sqrt{4\pi/(2J+1)}Y^{J*}_\lambda (\theta,\phi)$~\cite{edmonds57}. The angular state $|\theta \phi, \chi>$  can be expanded in terms of spherical harmonics
\begin{equation}
|\theta \phi, \chi> = \sum \limits_{J=0}^{\infty} \sum \limits_{\lambda=-J}^J Y^{J*}_{\lambda}(\theta, \phi) |J \lambda, \chi>
\end{equation}
The angular expansion of the $S$-matrix amplitudes 
$S_{\chi,0\nu}(\theta \phi)=<\theta \phi,\chi|<p_c p_d|S|p_ap_b,0\nu>$
\begin{equation}
S_{\chi,0\nu}(\theta\phi)= \sum \limits_{J=0}^{\infty} \sum \limits_{\lambda=-J}^J Y^J_{\lambda}(\theta,\phi) S^J_{\lambda \chi,0\nu}
\end{equation}
defines partial wave $S$-matrix amplitudes
\begin{equation} 
S^J_{\lambda \chi, 0 \nu}(p_cp_d,p_ap_b)=<J\lambda,\chi|<p_c p_d|S|p_ap_b,0\nu>
\end{equation}
of definite dipion spin. With $S_{\chi, 0 \nu}=i(2 \pi)^4 \delta^4 (P_f-P_i)T_{\chi, 0 \nu}$ the measured helicity amplitudes are defined by
\begin{equation}
H_{\chi,0 \nu}(s,t,m,\theta \phi)=\sqrt {q(m^2)G(s)/Flux(s)} T_{\chi,0\nu}(s,t,m,\theta\phi)
\end{equation}
where $s$ is the center-of-mass energy squared, $t=(p_c-p_a)^2$ is the four-momentum transfer squared, $q(m^2)G(s)$ is the final state Lorentz invariant phase space~\cite{svec97a} and $Flux(s)$ is the incident particles flux. The angular expansion of the production amplitudes (5.5) follows from (5.3)
\begin{equation}
H_{\chi,0\nu}(\theta\phi)= \sum \limits_{J=0}^{\infty} \sum \limits_{\lambda=-J}^J Y^J_{\lambda}(\theta,\phi) H^J_{\lambda \chi,0\nu}
\end{equation}
\noindent 
where $H^J_{\lambda \chi, 0 \nu}(s,t,m)$ are partial wave helicity amplitudes 
of definite dipion spin. 

\subsection{Constraints from conservation of $P$-parity}

Since pion helicities $\mu_1=\mu_2=0$ the two-pion angular states 
$|m,\vec{p_c}>\otimes|pJ\lambda;00>$ have a character of single particle helicity states for any fixed invariant mass $m$. The initial and final states in these processes are both separable. The helicity amplitudes $H^J_{\lambda \chi,0\nu}(s,t,m)$ then describe two-body scattering processes $\pi^- + p \to "J(m^2)" + n$ where $"J(m^2)"$ is the dipion "particle" with spin $J$ and mass $m$. Describing strong interactions, these amplitudes are expected to conserve $P$-parity.

$P$-parity conservation in strong interactions imposes constraints on two-body helicity amplitudes~\cite{martin70,perl74,leader01}
\begin{equation}
H_{-\mu_c-\mu_d,-\mu_a -\mu_b}=\eta(-1)^{\mu'-\mu}H_{\mu_c \mu_d,\mu_a\mu_b}
\end{equation}
where $\eta=\eta_a\eta_b\eta_c\eta_d(-1)^{s_c+s_d-s_a-s_b}$ and $\mu=\mu_a-\mu_b$,  $\mu'=\mu_c-\mu_d$. The derivation of these constraints requires that the initial and final states are both separable states of single particle helicity states and that the total angular momentum is conserved in the reaction~\cite{martin70}. These conditions are satisfied by the processes $\pi^- + p \to "J(m^2)" + n$ and the helicity amplitudes $H^J_{\lambda \chi,0\nu}$ thus must satisfy the parity constraints (5.28). From (4.4) we find that $\eta_{\pi\pi}=(-1)^J$. The parity constraints for $H^J_{\lambda \chi,0\nu}$ then read
\begin{equation}
H^J_{-\lambda - \chi,0 - \nu} = (-1)^{\lambda+ \chi + \nu} H^J_{\lambda \chi, 0 \nu}
\end{equation}
These parity constraints apply to all pion production processes $\pi N \to \pi \pi N$. The target nucleon and recoil nucleon helicities $\nu$ and $\chi$ are defined in the $s$-channel helicity system. The dipion helicity $\lambda$ will be defined in the $t$-channel helicity system~\cite{martin70,lutz78,becker79a}.

Assuming that all pion isospin states behave as identical particles in strong interactions the generalized Bose-Einstein statistics requires $I+J=even$ where $I$ is the dipion isospin~\cite{martin70}. As the result, the partial wave helicity amplitudes with odd spins $J$ vanish in reactions with two identical pions in the final state and the $\pi^- \pi^+$ isospin states are maximally entangled symmetric and antisymmetric Bell states for even and odd isispin, respectively.

\subsection{Nucleon helicity and transversity amplitudes with definite $t$-channel naturality}

The helicity amplitudes $H^J_{\lambda \chi,0 \nu}$ are combinations of helicity amplitudes with definite $t$-channel naturality $\eta = \mathcal{P} \mathcal{S}$
where $\mathcal{P}$ and $\mathcal{S}$ are the parity and the signature of Reggeons exchanged in $\pi^- + p \to "J(m^2)" + n$~\cite{martin70}. The natural and unnatural exchange amplitudes $N^J_{\lambda +,0 \pm}$ and $U^J_{\lambda +,0 \pm}$  correspond to naturality $\eta = +1$ and $\eta = -1$, respectively. They are given for $\lambda \neq 0$ by relations~\cite{lutz78,bourrely80,leader01}
\begin{eqnarray}
U^J_{\lambda +,0 \pm} & = & {1 \over{\sqrt{2}}} (H^J_{\lambda +,0 \pm} + 
(-1)^\lambda H^J_{- \lambda +,0 \pm})\\
\nonumber
N^J_{\lambda +,0 \pm} & = & {1 \over{\sqrt{2}}} (H^J_{\lambda +,0 \pm} - 
(-1)^\lambda H^J_{- \lambda +,0 \pm}) 
\end{eqnarray}
For $\lambda = 0$ they are 
\begin{equation}
U^J_{0+,0 \pm} = H^J_{0+,0 \pm}, \quad N^J_{0+,0 \pm} = 0
\end{equation}  
In (5.9) and (5.10) + and - correspond to $+{1 \over {2}}$ and $-{1 \over {2}}$ values of nucleon helicities. The unnatural exchange amplitudes $U^J_{\lambda +,0-}$ and $U^J_{\lambda +,0+}$ exchange $\pi$ and $a_1$ quantum numbers in the $t$-channel, respectively, while the natural exchange amplitudes $N^J_{\lambda +,0-}$ and $N^J_{\lambda +,0+}$ both exchange $a_2$ quantum numbers.

Amplitude analyses of measurements on polarized targets are best performed in terms of transversity amplitudes with definite $t$-channel naturality~\cite{lutz78}. In such measurements the spin states of the target nucleon are described by transversity $\tau$ with $\tau = +{1 \over{2}} \equiv u$ and $\tau = -{1 \over{2}} \equiv d$ corresponding to "up" and "down" orientations of the target spin relative to the scattering plane~\cite{bourrely80,leader01}. Following Lutz and Rybicki~\cite{lutz78}, we define mixed helicity-transversity amplitudes with nucleon helicity replaced by nucleon transversity
\begin{equation}
T^J_{\lambda \tau_n,0 \tau} = \sum\limits_{\chi,\nu} D^{{1\over{2}}*}_{\tau_n \chi }({\pi \over{2}}, {\pi \over{2}},-{\pi \over{2}})e^{i \pi (\chi-\nu)}H^J_{\lambda \chi,0 \nu}D^{1\over{2}}_{\nu \tau}({\pi \over{2}}, {\pi \over{2}},-{\pi \over{2}})
\end{equation}
Using the parity relations (5.8) for helicity amplitudes we obtain
\begin{eqnarray}
T^J_{\lambda u,0u} & = & {1 \over{2}}\bigl(1-(-1)^\lambda \bigr) \Bigl(H^J_{\lambda +,0+} + iH^J_{\lambda +,0-}\Bigr)\\
\nonumber
T^J_{\lambda d,0d} & = & {1 \over{2}}\bigl(1-(-1)^\lambda \bigr)\Bigl(H^J_{\lambda +,0+} - iH^J_{\lambda +,0-}\Bigr)\\
\nonumber
-iT^J_{\lambda d,0u} & = & {1 \over{2}}\bigl(1+(-1)^\lambda \bigr)
\Bigl(H^J_{\lambda +,0+} + iH^J_{\lambda+,0-}\Bigr)\\
\nonumber
+iT^J_{\lambda u,0d} & = & {1 \over{2}}\bigl(1+(-1)^\lambda \bigr)
\Bigl(H^J_{\lambda +,0+} - iH^J_{\lambda +,0-}\Bigr)
\end{eqnarray}
As a result of parity conservation the following amplitudes vanish
\begin{eqnarray}
T^J_{\lambda \tau,0\tau} & = 0 \quad  \lambda \text{=even}\\
\nonumber
T^J_{\lambda -\tau,0\tau} & = 0 \quad  \lambda \text{= odd}
\end{eqnarray}
Absorbing the inessential factors $\pm i$ in front of $T^J_{\lambda d,0u}$ and $T^J_{\lambda u,0d}$ in (5.12) into these amplitudes, the unnatural and natural exchange transversity amplitudes are given for $\lambda = even$ by
\begin{eqnarray}
U^J_{\lambda,\tau} & = &
{1 \over{2}} (T^J_{\lambda -\tau,0 \tau}+(-1)^\lambda T^J_{-\lambda -\tau,0 \tau})\\
\nonumber 
N^J_{\lambda,\tau} & = &  
{1 \over{2}} (T^J_{\lambda -\tau,0 \tau}-(-1)^\lambda T^J_{- \lambda -\tau,0 \tau})
\end{eqnarray}
and for $\lambda = odd$ by
\begin{eqnarray}
U^J_{\lambda,\tau} & = &
{1 \over{2}} (T^J_{\lambda \tau,0 \tau}+(-1)^\lambda T^J_{- \lambda \tau,0 \tau})\\
\nonumber 
N^J_{\lambda,\tau} & = &  
{1 \over{2}} (T^J_{\lambda \tau,0 \tau}-(-1)^\lambda T^J_{- \lambda \tau,0 \tau})
\end{eqnarray}
For $\lambda= even$ the recoil nucleon transversity $\tau_n=-\tau$. For $\lambda= odd$, $\tau_n=+\tau$. Note that $N^J_{0,\tau}=0$. From (5.14) and (5.15) we find parity relations
\begin{equation}
U^J_{-\lambda, \tau}=+(-1)^\lambda U^J_{\lambda, \tau}, \quad N^J_{-\lambda, \tau}=-(-1)^\lambda N^J_{\lambda, \tau}
\end{equation}

It is useful to express transversity amplitudes $U^J_{\lambda, \tau}$ and $N^J_{\lambda, \tau}$ in terms of unnatural and natural exchange helicity amplitudes (5.9). Using (5.12) and (5.9) in (5.14) and (5.15) we find 
\begin{eqnarray}
U^J_{\lambda,u} & = & {1 \over{\sqrt{2}}} (U^J_{\lambda +,0+} + i U^J_{\lambda +,0-})\\
\nonumber
U^J_{\lambda,d} & = & {1 \over{\sqrt{2}}} (U^J_{\lambda +,0+} - i U^J_{\lambda +,0-})\\
N^J_{\lambda,u} & = & {1 \over{\sqrt{2}}} (N^J_{\lambda +,0+} + i N^J_{\lambda +,0-})\\
\nonumber
N^J_{\lambda,d} & = & {1 \over{\sqrt{2}}} (N^J_{\lambda +,0+} - i N^J_{\lambda +,0-})
\end{eqnarray}

\section{Final state density matrix in $\pi N \to \pi \pi N$ processes.}

\subsection{Angular final state density matrix}

The pion beam and nucleon target are prepared in an initial state $\rho_i = \rho_i(\pi_a) \otimes \rho_i(N_b,\vec{P})$ where $\rho_i(\pi_a) = |p_a0 > <p_a0|$ and
\begin{equation}
\rho_i(N_b,\vec{P})=\sum \limits_{\nu \nu^{'}} \rho_b(\vec{P})_{\nu \nu^{'}} |p_b \nu> <p_b \nu^{'}|
\end{equation}
$\rho_b (\vec{P})$ is the target nucleon spin density matrix 
\begin{equation}
\rho_b (\vec{P})={1\over{2}}(1+\vec{P}\vec{\sigma})
\end{equation}
where $\vec{P}=(P_x,P_y,P_z)$ is the target polarization vector, $\vec{\sigma} = (\sigma_x, \sigma_y, \sigma_z)$ are Pauli matrices and $Tr(\rho_b(\vec{P}))=1$. Following (3.5) the density matrix $\rho_f(\vec{P})$ of the final state $\pi_1 \pi_2 N_d$ reads
\begin{equation} 
\rho_f(\vec{P})=
\sum \limits_{\chi \chi'} \int d\Phi_3 d\Phi_3'
<\theta \phi,\chi,p_c p_d|S \rho_i S^+|p_c' p_d',\theta' \phi',\chi'>
|p_cp_d,\theta \phi,\chi><\theta' \phi',\chi',p_c'p_d'|
\end{equation}
where we have used the phase space relation~\cite{martin70} 
\begin{equation}
d\Phi_3=
{d^3 \vec{p}_1 \over{2E_1}}{d^3 \vec{p}_2 \over{2E_2}}{d^3 \vec{p}_3 \over{2E_3}}
={q\over{4m}}d^4p_c d\Omega{d^3 \vec{p}_3 \over{2E_3}}=
d\overline{\Phi}_3 d\Omega
\end{equation}
in the completness relation (3.2). We shall use the projection of $\rho_f(\vec{P})$ into an angular state $\rho_f(p_cp_d,\theta \phi, \vec{P})$ with definite final state momenta
\begin{equation}
\rho_f(p_cp_d,\theta \phi, \vec{P})
=\sum \limits_{\chi \chi^{'}} 
\rho_f(p_cp_d,\theta \phi, \vec{P})_{\chi \chi{'}} |\chi><\chi^{'}|
\end{equation}
In the following we suppress the momentum labels in the initial and final helicity states. The density matrix elements are given by the $S$-matrix evolution law
\begin{equation}
\rho_f(\theta \phi, \vec{P})_{\chi \chi'}=\sum \limits_{\nu \nu'} 
<\theta \phi,\chi|S|0 \nu>\rho_b (\vec{P})_{\nu \nu'}<0 \nu'|S^+|\theta \phi,\chi'>
\end{equation}  
With $S_{\chi, 0 \nu}=<\theta \phi, \chi|S|0 \nu>=i(2 \pi)^4 \delta^4 (P_f-P_i)T_{\chi, 0 \nu}$ we get
\begin{equation}
\rho_f(\theta \phi, \vec{P})_{\chi \chi'}=\rho^{'}_f(\theta \phi, \vec{P})_{\chi \chi'}(VT) (2\pi)^4 \delta^4(P_f-P_i)
\end{equation}
\noindent
where $\rho^{'}_f(\theta \phi, \vec{P})$ is expressed in terms of transition amplitudes $T_{\chi, 0 \nu}$ and where we have used the conventional approach to deal with a square of $\delta$-functions~\cite{perl74} with $V$ and $T$ being total volume and time confining the interactions to be taken in the limit $V,T \to \infty$. According to the Born rule, the probability of $\pi_a N_b(\nu) \to \pi_1 \pi_2 N_d(\chi)$ is given by
\begin{equation}
dP_{\chi,0\nu}=|S_{\chi,0\nu}|^2 \prod \limits^d_{n=1} {d^3 \vec{p}_n \over{2E_n}}=
|T_{\chi,0\nu}|^2 dLips(P_i,p_1,p_2,p_d)(VT)
\end{equation}
\noindent
Here $P_i=p_a+p_b$ is the total four-momentum and the Lorentz invariant phase space $dLips=q(m^2)G(s)dmdtd\Omega$ where $G(s)$ is energy dependent part of the phase space~\cite{svec97a}. The probability per unit volume, unit time and per target particle is $d\sigma_{\chi,0\nu}=dP_{\chi,0\nu}/(VTFlux(s))$ and the differential cross-section reads
\begin{equation}
{d\sigma_{\chi,0\nu} \over {dtdmd\Omega}}={q(m^2)G(s) \over {Flux(s)}}|T_{\chi,0\nu}|^2
\end{equation}
\noindent
Applying formally the same procedure to every bilinear term $S_{\chi,0\nu}S_{\chi',0\nu'}^*$ of $\rho_f(\theta \phi ,\vec{P})$ we can define a differential cross-section matrix 
\begin{equation}
{d\sigma \over {dtdmd\Omega}}={q(m^2)G(s) \over {Flux(s)}}\rho^{'}_f(\theta \phi,\vec{P}) \equiv \rho_f(\theta \phi ,\vec{P})
\end{equation}
\noindent
where we have redefined the final state density matrix $\rho_f(\theta \phi ,\vec{P})$ to read
\begin{equation}
\rho_f(\theta \phi, \vec{P})_{\chi \chi'}=\sum \limits_{\nu \nu'} 
H_{\chi,0 \nu}(\theta \phi)\rho_b (\vec{P})_{\nu \nu'}H_{\chi',0 \nu'}(\theta \phi)
\end{equation}  
The redefined transition amplitudes $H_{\chi,0 \nu}(s,t,m,\theta \phi)$ are
given by (5.5) in Section V.
 
\subsection{Recoil nucleon polarization}

To discuss the structure of the angular final state density matrix (6.12) we first note a useful result from quantum state tomography~\cite{nielsen00}. Arbitrary density matrix $\rho$ of $n$ qubits can be expanded in a form 
\begin{equation}
\rho = \sum \limits_{\vec{v}} ({1 \over{2^n}}) Tr(\sigma_{v_1} \otimes \sigma_{v_2} \otimes ... \otimes \sigma_{v_n} \rho) \sigma_{v_1} \otimes \sigma_{v_2} \otimes ... \otimes \sigma_{v_n}
\end{equation}
\noindent
where the sum is over the vectors $\vec{v} = (v_1, v_2, ..., v_n)$ with entries chosen from the set $\sigma^j, j=0,1,2,3$ of Pauli matrices and $\sigma^0 = \openone$. The traces in (6.12) represent average values of spin correlations. The final density matrix $\rho_f(\theta \phi, \vec{P})$ is a single qubit density matrix corresponding to spin ${1 \over{2}}$ of the recoil nucleon. It can be written in the form (6.12)
\begin{equation}
\rho_f(\theta \phi, \vec{P}) = {1 \over{2}} \bigl (I^0 (\theta \phi, \vec{P}) \sigma^0 + {\vec{I}} (\theta \phi, \vec{P}) \vec{\sigma} \bigr )
\end{equation}
\noindent
where the traces $I^j(\theta \phi, \vec{P}) = Tr(\sigma^j \rho_f(\theta \phi, \vec{P})), j=0,1,2,3$ represent measurable intensities of angular distributions. The density matrix can be written in an equivalent form in the recoil nucleon helicity basis $|\chi><\chi'|$
\begin{equation}
\rho_f(\theta \phi, \vec{P}) =
\left( \begin{array}{cc}
I^0 (\theta \phi, \vec{P})+ I^3 (\theta \phi, \vec{P}), & 
I^1 (\theta \phi, \vec{P})-iI^2 (\theta \phi, \vec{P})\\
I^1 (\theta \phi, \vec{P})+iI^2 (\theta \phi, \vec{P}), &
I^0 (\theta \phi, \vec{P})- I^3 (\theta \phi, \vec{P})\\
\end{array} \right)
\end{equation}
Introducing recoil nucleon polarization vector $\vec{Q} (\theta \phi, \vec{P})$ using a relation~\cite{leader01,bourrely80}
\begin{equation}
\vec{I} (\theta \phi, \vec{P}) \equiv \vec{Q} (\theta \phi, \vec{P}) I^0(\theta \phi, \vec{P})  
\end{equation}
we can write 
\begin{equation}
\rho_f(\theta \phi, \vec{P}) = {1 \over{2}} \bigl (1+\vec{Q}(\theta \phi, \vec{P}) \vec{\sigma}\bigr ) I^0 (\theta \phi, \vec{P})= \rho_d(\vec {Q}) I^0 (\theta \phi, \vec{P}) 
\end{equation}
where $\rho_d(\vec {Q})={1 \over{2}} (1+\vec{Q} \vec{\sigma})$ is the recoil nucleon spin density matrix. The polarization vector $\vec{Q}=(Q^1,Q^2,Q^3)$ is defined in the rest frame of the recoil nucleon. It has transverse component $Q^2$ that is perpendicular to the scattering plane in the direction of the $y$ axis. The transverse component $Q^1$ is perpendicular to the direction of recoil nucleon in the scattering plane. The longitudinal component $Q^3$ is along the direction of its motion.

The equation (6.16) is our principal result. It shows that the purity of the final state $\rho_f(\theta \phi, \vec{P})$ is controlled entirely by the recoil nucleon polarization $\vec{Q}(\theta \phi, \vec{P})$ which we shall relate in the following to partial wave transversity amplitudes under the assumption of $P$-parity conservation. 

\subsection{Angular intensities in terms of density matrix elements}

To measure the final state density matrix (6.13) requires measurements of angular intensities the measurement of which is further reduced to the measurements of density matrix elements in terms of amplitudes. Using  the target nucleon spin density matrix (6.2) in the expression (6.11) we can write the matrix elements of $\rho_f(\theta \phi, \vec{P})$ in terms of components of target polarization
\begin{equation}
\rho_f(\theta \phi, \vec{P})_{\chi \chi'}= 
\rho_u(\theta \phi)_{\chi\chi^{'}}+P_x\rho_x(\theta \phi)_{\chi\chi^{'}}+
P_y\rho_y(\theta \phi)_{\chi \chi^{'}}+P_z\rho_z(\theta \phi)_{\chi \chi^{'}}
\end{equation}
where the subscript $u$ stands for unpolarized target $\vec{P} = 0$. The polarization components of density matrix elements in (6.17) are given by (6.11)
\begin{equation} 
\rho_k(\theta\phi)_{\chi\chi^{'}} = {1 \over{2}}\sum \limits_{\nu\nu^{'}}H_{\chi,0\nu}(\theta\phi)(\sigma_k)_{\nu\nu^{'}}H^*_{\chi^{'},0\nu^{'}}
(\theta\phi)
\end{equation}
\noindent
where $k=u,x,y,z$ and $\sigma_u \equiv \sigma_0$. Using (5.6) for $H_{\chi,0\nu}(\theta\phi)$ their angular expansion reads
\begin{equation}
\rho_k(\theta \phi)_{\chi \chi^{'}}= \sum \limits_{J \lambda} \sum  \limits_{J^{'} \lambda^{'}} (R_k)^{J {1 \over{2}}, J^{'} {1 \over{2}}}_{\lambda \chi, \lambda^{'} \chi^{'}} Y^{J}_{\lambda}(\theta,\phi) Y^{J^{'*}}_{\lambda^{'}}(\theta, \phi)
\end{equation}
\noindent
where 
\begin{equation}
(R_k)^{J{1 \over{2}}, J^{'} {1 \over{2}}}_{\lambda\chi, \lambda^{'} \chi^{'}}=
{1 \over{2}} \sum \limits_{\nu \nu^{'}} H^J_{\lambda \chi ,0 \nu} (\sigma_k)_{\nu \nu^{'}} H^{J^{'}*}_{\lambda^{'} \chi^{'},0 \nu^{'}}
\end{equation}
Using the decomposition (6.17) for $\rho_f(\theta \phi, \vec{P})$ we find a decomposition for the intensities $I^j(\theta \phi, \vec{P})$ in (6.13) in terms of components of the target polarization
\begin{equation}
I^j(\theta \phi, \vec{P})= Tr(\sigma^j \rho_f(\theta \phi,\vec{P})) =
I^j_u(\theta \phi) + P_x I^j_x(\theta \phi)+ P_y I^j_y(\theta \phi)+ P_z I^j_z(\theta \phi)
\end{equation}
where the components $I^j_k(\theta \phi)$, $j=0,1,2,3$ and $k=u,x,y,z$ are given by traces
\begin{equation}
I^j_k(\theta \phi) = Tr_{\chi, \chi'} \bigl ((\sigma^j)_{\chi' \chi} 
\rho_k(\theta \phi)_{\chi \chi'}\bigr )
\end{equation}
The component intensities $I^j_k(\theta \phi)$ have angular expansions arising from these traces
\begin{equation}
I^j_k(\theta \phi) =\sum \limits_{J \lambda} \sum \limits_{J' \lambda'} (R^j_k)^{JJ'}_{\lambda \lambda'} Y^J_{\lambda}(\theta, \phi) Y^{J'*}_{\lambda'}(\theta \phi)
\end{equation}
\noindent
where the unnormalized dipion density matrix elements $(R^j_k)^{JJ'}_{\lambda \lambda'}$ are  traces over recoil nucleon helicities
\begin{equation}
(R^j_k)^{JJ'}_{\lambda \lambda'}= Tr_{\chi, \chi'}\bigl ((\sigma^j)_{\chi' \chi}(R_k)^{J{1 \over{2}}, J^{'} {1 \over{2}}}_{\lambda\chi, \lambda^{'} \chi'}\bigr )
\end{equation}
\noindent 
Expressed in terms of partial wave helicity amplitudes they read
\begin{equation}
(R^j_k)^{JJ'}_{\lambda \lambda'} = {1 \over{2}} \sum \limits_{\chi \chi'} \sum \limits_{\nu \nu'} (\sigma^j)_{\chi' \chi}H^J_{\lambda \chi, 0 \nu}(\sigma_k)_{\nu \nu'} H^{J'*}_{\lambda' \chi', 0 \nu'}
\end{equation}
Combining (6.19) and (6.23) in (6.14) we can express the density matrix elements (6.20) in terms of density density matrix elements (6.25) for each $k=u,x,y,z$
\begin{eqnarray}
(R_k)^{J{1 \over{2}}, J^{'} {1 \over{2}}}_{\lambda +, \lambda^{'} +} & = &
(R^0_k)^{JJ'}_{\lambda \lambda'}+(R^3_k)^{JJ'}_{\lambda \lambda'}\\
\nonumber
(R_k)^{J{1 \over{2}}, J^{'} {1 \over{2}}}_{\lambda +, \lambda^{'} -} & = & 
(R^1_k)^{JJ'}_{\lambda \lambda'}-i(R^2_k)^{JJ'}_{\lambda \lambda'}\\
\nonumber
(R_k)^{J{1 \over{2}}, J^{'} {1 \over{2}}}_{\lambda -, \lambda^{'} +} & = & 
(R^1_k)^{JJ'}_{\lambda \lambda'}+i(R^2_k)^{JJ'}_{\lambda \lambda'}\\
\nonumber
(R_k)^{J{1 \over{2}}, J^{'} {1 \over{2}}}_{\lambda -, \lambda^{'} -} & = & 
(R^0_k)^{JJ'}_{\lambda \lambda'}-(R^3_k)^{JJ'}_{\lambda \lambda'}
\end{eqnarray}

\subsection{Constraints on angular intensities from $P$-parity conservation}

Conservation of $P$-parity in strong interactions is encoded in the angular expansion of angular final state density matrix by imposing parity relations on the partial wave helicity amplitudes. Relabeling the summations in (6.23) and combining in the sum (6.23) the terms with inverted $J \lambda$ and $J' \lambda'$ we can write the sum (6.23) for each $k=u,y,x,y$ and $j=0,1,2,3$ as the sum of four terms
\begin{equation}
{1 \over{4}} \sum \limits_{J \lambda} \sum \limits_{J' \lambda'} 
[R^{JJ'}_{\lambda \lambda'} Y^J_{\lambda} Y^{J'*}_{\lambda'} + 
R^{J'J}_{\lambda' \lambda} Y^{J'}_{\lambda'} Y^{J*}_{\lambda} + 
R^{JJ'}_{-\lambda -\lambda'} Y^J_{-\lambda} Y^{J'*}_{-\lambda'} + 
R^{J'J}_{-\lambda' -\lambda} Y^{J'}_{-\lambda'} Y^{J*}_{-\lambda}]
\end{equation}
\noindent
Using hermiticity of the density matrix
\begin{equation}
(R^j_k)^{J'J}_{\lambda' \lambda}=(R^j_k)^{JJ'*}_{\lambda \lambda'}
\end{equation}
\noindent
and a relation for spherical harmonics $Y^L_{-M} (\theta, \phi) = (-1)^M (Y^L_{M} (\theta, \phi))^{*}$ the sum of terms in (6.27) takes the form
\begin{equation} 
[+2Re(R^{JJ'}_{\lambda \lambda'} + (-1)^{\lambda+ \lambda'} R^{J J' *}_{- \lambda -\lambda'})
Re(Y^J_\lambda Y^{J'*}_{\lambda'})
\end{equation}
\[
 -2Im(R^{JJ'}_{\lambda \lambda'} - (-1)^{\lambda+ \lambda'} R^{J J' *}_{- \lambda -\lambda'})
Im(Y^J_\lambda Y^{J'*}_{\lambda'})]
\]
Parity relations (5.8) for the partial wave helicity amplitudes 
$H^J_{\lambda \chi ,0 \nu}$ imply parity relations for the density matrix elements
\begin{equation}
(R^j_k)^{JJ'}_{\lambda \lambda'} = + (-1)^{\lambda + \lambda'} (R^j_k)^{JJ'}_{-\lambda -\lambda'} 
\end{equation}
\noindent

\begin{table}
\caption{Density matrix elements expressed in terms of nucleon transversity amplitudes with definite $t$-channel naturality. The spin indices $JJ'$ which always go with helicities $\lambda \lambda'$ have been omitted in the amplitudes. The coefficients $\eta_\lambda=1$ for $\lambda=0$ and $\eta_\lambda=1/\sqrt{2}$ for $\lambda \neq 0$.}
\begin{tabular}{l|r}
\toprule
$(R^0_u)^{JJ'}_{\lambda \lambda'}$ & $\eta_\lambda \eta_{\lambda'} [U_{\lambda,u}U^*_{\lambda',u}+N_{\lambda,u}N^*_{\lambda',u}
+U_{\lambda,d}U^*_{\lambda',d}+N_{\lambda,d}N^*_{\lambda',d}]$\\
$(R^0_y)^{JJ'}_{\lambda \lambda'}$ & $\eta_\lambda \eta_{\lambda'} [U_{\lambda,u}U^*_{\lambda',u}+N_{\lambda,u}N^*_{\lambda',u}
-U_{\lambda,d}U^*_{\lambda',d}-N_{\lambda,d}N^*_{\lambda',d}]$\\
$(R^0_x)^{JJ'}_{\lambda \lambda'}$ & $-i\eta_\lambda \eta_{\lambda'} [U_{\lambda,u}N^*_{\lambda',d}+N_{\lambda,u}U^*_{\lambda',d}
-U_{\lambda,d}N^*_{\lambda',u}-N_{\lambda,d}U^*_{\lambda',u}]$\\
$(R^0_z)^{JJ'}_{\lambda \lambda'}$ & $\eta_\lambda \eta_{\lambda'} [U_{\lambda,u}N^*_{\lambda',d}+N_{\lambda,u}U^*_{\lambda',d}
+U_{\lambda,d}N^*_{\lambda',u}+N_{\lambda,d}U^*_{\lambda',u}]$\\
\colrule

$(R^2_u)^{JJ'}_{\lambda \lambda'}$ & $-\eta_\lambda \eta_{\lambda'} [U_{\lambda,u}U^*_{\lambda',u}-N_{\lambda,u}N^*_{\lambda',u}
-U_{\lambda,d}U^*_{\lambda',d}+N_{\lambda,d}N^*_{\lambda',d}]$\\
$(R^2_y)^{JJ'}_{\lambda \lambda'}$ & $-\eta_\lambda \eta_{\lambda'} [U_{\lambda,u}U^*_{\lambda',u}-N_{\lambda,u}N^*_{\lambda',u}
+U_{\lambda,d}U^*_{\lambda',d}-N_{\lambda,d}N^*_{\lambda',d}]$\\
$(R^2_x)^{JJ'}_{\lambda \lambda'}$ & $i\eta_\lambda \eta_{\lambda'} [U_{\lambda,u}N^*_{\lambda',d}-N_{\lambda,u}U^*_{\lambda',d}
+U_{\lambda,d}N^*_{\lambda',u}-N_{\lambda,d}U^*_{\lambda',u}]$\\
$(R^2_z)^{JJ'}_{\lambda \lambda'}$ & $-\eta_\lambda \eta_{\lambda'} [U_{\lambda,u}N^*_{\lambda',d}-N_{\lambda,u}U^*_{\lambda',d}
-U_{\lambda,d}N^*_{\lambda',u}+N_{\lambda,d}U^*_{\lambda',u}]$\\
\colrule

$(R^1_u)^{JJ'}_{\lambda \lambda'}$ & $-i\eta_\lambda \eta_{\lambda'} [U_{\lambda,u}N^*_{\lambda',u}-N_{\lambda,u}U^*_{\lambda',u}
-U_{\lambda,d}N^*_{\lambda',d}+N_{\lambda,d}U^*_{\lambda',d}]$\\
$(R^1_y)^{JJ'}_{\lambda \lambda'}$ & $-i\eta_\lambda \eta_{\lambda'} [U_{\lambda,u}N^*_{\lambda',u}-N_{\lambda,u}U^*_{\lambda',u}
+U_{\lambda,d}N^*_{\lambda',d}-N_{\lambda,d}U^*_{\lambda',d}]$\\
$(R^1_x)^{JJ'}_{\lambda \lambda'}$ & $-\eta_\lambda \eta_{\lambda'} [U_{\lambda,u}U^*_{\lambda',d}-N_{\lambda,u}N^*_{\lambda',d}
+U_{\lambda,d}U^*_{\lambda',u}-N_{\lambda,d}N^*_{\lambda',u}]$\\
$(R^1_z)^{JJ'}_{\lambda \lambda'}$ & $-i\eta_\lambda \eta_{\lambda'} [U_{\lambda,u}U^*_{\lambda',d}-N_{\lambda,u}N^*_{\lambda',d}
-U_{\lambda,d}U^*_{\lambda',u}+N_{\lambda,d}N^*_{\lambda',u}]$\\
\colrule

$(R^3_u)^{JJ'}_{\lambda \lambda'}$ & $\eta_\lambda \eta_{\lambda'} [U_{\lambda,u}N^*_{\lambda',u}+N_{\lambda,u}U^*_{\lambda',u}
+U_{\lambda,d}N^*_{\lambda',d}+N_{\lambda,d}U^*_{\lambda',d}]$\\
$(R^3_y)^{JJ'}_{\lambda \lambda'}$ & $\eta_\lambda \eta_{\lambda'} [U_{\lambda,u}N^*_{\lambda',u}+N_{\lambda,u}U^*_{\lambda',u}
-U_{\lambda,d}N^*_{\lambda',d}-N_{\lambda,d}U^*_{\lambda',d}]$\\
$(R^3_x)^{JJ'}_{\lambda \lambda'}$ & $-i\eta_\lambda \eta_{\lambda'} [U_{\lambda,u}U^*_{\lambda',d}+N_{\lambda,u}N^*_{\lambda',d}
-U_{\lambda,d}U^*_{\lambda',u}-N_{\lambda,d}N^*_{\lambda',u}]$\\
$(R^3_z)^{JJ'}_{\lambda \lambda'}$ & $\eta_\lambda \eta_{\lambda'} [U_{\lambda,u}U^*_{\lambda',d}+N_{\lambda,u}N^*_{\lambda',d}
+U_{\lambda,d}U^*_{\lambda',u}+N_{\lambda,d}N^*_{\lambda',u}]$\\

\botrule
\end{tabular}
\label{Table I.}
\end{table}

for $(k,j)=(u,0),(y,0),(u,2),(y,2),(x,1),(z,1),(x,3),(z,3)$ and
\begin{equation}
(R^j_k)^{JJ'}_{\lambda \lambda'} = - (-1)^{\lambda + \lambda'} (R^j_k)^{JJ'}_{-\lambda -\lambda'} 
\end{equation}
\noindent
for $(x,0),(z,0),(x,2),(z,2),(u,1),(y,1),(u,3),(y,3)$. Using these symmetry relations in (6.29) the components $I^j_k(\theta \phi)$ of the dipion angular distribution $I^j(\theta \phi, \vec{P})$ measured on polarized target take the form
\begin{equation}
I^j_k(\theta \phi)=\sum \limits_{J \lambda} \sum \limits_{J' \lambda'} 
(Re R^j_k)^{JJ'}_{\lambda \lambda'} Re(Y^J_\lambda(\theta \phi)Y^{J'*}_{\lambda'}(\theta \phi))
\end{equation}
\noindent
for $(k,j)=(u,0),(y,0),(u,2),(y,2),(x,1),(z,1),(x,3),(z,3)$ and
\begin{equation}
I^j_k(\theta \phi)=-\sum \limits_{ J \lambda} \sum \limits_{J' \lambda'} 
(Im R^j_k)^{JJ'}_{\lambda \lambda'} Im(Y^J_\lambda(\theta \phi)Y^{J'*}_{\lambda'}(\theta \phi))
\end{equation}
\noindent
for $(x,0),(z,0),(x,2),(z,2),(u,1),(y,1),(u,3),(y,3)$. The elements 
$(Im R^j_k)^{JJ'}_{\lambda \lambda'}$ in the group (6.32) and $(Re R^j_k)^{JJ'}_{\lambda \lambda'}$ in the group (6.33) are not observable as the result of parity conservation.

Because of the angular properties of $Y^1_\lambda(\theta \phi)$, the three elements $(R^j_k)^{00}_{00}$, $(R^j_k)^{11}_{00}$ and $(R^j_k)^{11}_{11}$ in (6.32) are not independent in $\pi^- p \to \pi^-\pi^+ n$ but appear in two independent combinations 
\begin{equation}
(R^j_k)_{SP} \equiv (R^j_k)^{00}_{00}+(R^j_k)^{11}_{00}+2(R^j_k)^{11}_{11}, 
\qquad
(R^j_k)_{PP} \equiv (R^j_k)^{11}_{00}-(R^j_k)^{11}_{11}
\end{equation}
What is usually measured in actual experiments are normalized density matrix elements $(\rho^j_k)^{JJ'}_{\lambda \lambda'}$ defined by
\begin{equation}
(R^j_k)^{JJ'}_{\lambda \lambda'}={d^2 \sigma \over{dtdm}}(\rho^j_k)^{JJ'}_{\lambda \lambda'} 
\end {equation}
where
\begin{equation}
{d^2 \sigma \over{dtdm}} \equiv \int d\Omega I^0_u(\theta\phi)=
\sum \limits_{J \lambda}(R^0_u)^{JJ}_{\lambda \lambda} = 
{1 \over{2}} \sum \limits_{J \lambda} \sum \limits_{\chi, \nu} |H^J_{\lambda \chi,0 \nu}|^2
\end{equation}
is the integrated intensity of $\pi \pi$ production measured on unpolarized target.

We have expressed the density matrix elements $(R^j_k)^{JJ'}_{\lambda \lambda'}$ in terms of transversity amplitudes $U^J_{\lambda,\tau}$ and $N^J_{\lambda,\tau}$. The results are are given in the Table I. and agree with Ref.~\cite{lutz78}. The normalization of the amplitudes is given by the trace (6.36)
\begin{equation} 
{d^2 \sigma \over{dtdm}}=
\sum \limits_{J, \lambda \geq 0} \sum \limits_{\tau} |U^J_{\lambda,\tau}|^2 + |N^J_{\lambda, \tau}|^2
\end{equation}

\section{Unitary evolution constraints in $\pi N \to \pi \pi N$ processes.}

\subsection{Constraints on angular intensities}

The initial $\pi N$ state is in a pure state only when the target nucleon is in a pure spin state. The target nucleon spin density matrix has the form 
$\rho_b (\vec {P}) = {1 \over{2}} (1+\vec{P} \vec{\sigma})$
where $\vec{P}=(P_x,P_y,P_z)$ is the target polarization vector. The target is in a pure state if and only if $|\vec{P}|^2=1$~\cite{nielsen00} or, equivalently, $\det(\rho_b(\vec{P}))=1-P_x^2-P_y^2-P_z^2=0$. This condition can be written in the form
\begin{equation}
\sum \limits_{m,n=u}^z \eta_{mn}P_m P_n=0
\end{equation}
where $P_u=1$ and $\eta_{mn}= diag(+1,-1,-1,-1)$ is Minkowski metric. In modern polarized targets the density matrix $\rho_b(\vec {P})$ is a mixed state with $|\vec{P}|^2<1$ that can be varied by using external magnetic fields to rotate the polarization vector $\vec{P}$ into any desired direction~\cite{leader01}. For $\vec{P}=0$ the target is unpolarized. The pure states define a Bloch sphere within which are located the mixed states.

As shown in (6.16), the normalized final state density matrix is equal to the recoil nucleon spin density matrix $\rho_d(\vec{Q})={1 \over{2}} (1+\vec{Q} \vec{\sigma})$. The final $\pi \pi N$ state is therefore in a pure state if and only if $|\vec{Q}|^2=1$ or, in terms of the intensities (6.15), if and only if $(I^0)^2-(I^1)^2-(I^2)^2-(I^3)^2=0$. Using the decomposition (6.21) this last condition takes the form
\begin{equation}
\sum \limits_{m,n=u}^z A_{mn}P_mP_n=0
\end{equation}
where
\begin{equation}
A_{mn}=\sum \limits_{j=1}^3 I_m^j I_n^j -I_m^0 I_n^0
\end{equation}
The purity condition (7.2) must hold true for all polarization vectors $\vec{P}$ which satisfy the purity condition (7.1) and for all values of the kinematic variables $s,t,m,\theta,\phi$. The intensities $I^j_k$ do not depend on the components $P_m$ of the polarization vector because the $S$-matrix amplitudes do not depend on the target polarization vector. With all terms $A_{mn}$ independent of the polarization vector, the condition (7.2) is not an independent quadratic form in $P_mP_n$ on the entire Bloch sphere but must coincide with the condition (7.1). That happens if and only if $A_{mn}=\eta_{mn}Z(s,t,m,\theta \phi)$ which implies that $Z=A_{uu}$. Unitary evolution then imposes 9 independent constraints
\begin{equation}
A_{mn}=\eta_{mn}A_{uu}
\end{equation}
Explicitely, the unitary evolution constraints on angular intensities read
\begin{equation}
\sum \limits_{j=1}^3 (I_u^j)^2+(I_k^j)^2=(I_u^0)^2+(I_k^0)^2
\end{equation}
for $A_{kk}=-A_{uu}$, $k=x,y,z$,
\begin{equation}
\sum_{j=1}^3 I_u^jI_k^j=I_u^0I_k^0
\end{equation}
for $A_{uk}=0$, $k=x,y,z$ and
\begin{equation}
\sum \limits_{j=1}^3 I_m^jI_n^j=I_m^0I_n^0
\end{equation}
for $A_{mn}=0$, $m,n=x,y,z$ and $m \neq n$. The three constraints (7.5) are equivalent to three constraints
\begin{equation}
\sum \limits_{j=1}^3 (I_m^j)^2-(I_n^j)^2=(I_m^0)^2-(I_n^0)^2
\end{equation}
for $A_{mm}-A_{nn}=0$, $m,n=x,y,z$ and $m \neq n$. The constraints (7.5) and (7.6) can be combined to read
\begin{equation}
\sum \limits_{j=1}^3 (I_u^j \pm I_k^j)^2=(I_u^0 \pm I_k^0)^2
\end{equation}
for $k=x,y,z$. These last constraints (7.9) are identical to conditions $|\vec{Q}|^2=1$ for special pure initial states with polarizations $P_k= \pm1$.

Mixed target spin states evolve into mixed recoil nucleon spin states. The conditions (7.1) and (7.2) change to read
\begin{eqnarray}
\sum \limits_{m,n=u}^z \eta_{mn}P_m P_n & > & 0 \\
\sum \limits_{m,n=u}^z    A_{mn}P_m P_n & > & 0
\end{eqnarray}
implying the same constraints (7.4) as in the case of the pure states. The conditions (7.11) also exclude the possibility that $A_{mn}=0$ for all $m,n$ which is allowed by (7.2).

\subsection{Constraints on parity conserving nucleon transversity amplitudes}

We looked for constraints on parity conserving transversity amplitudes implied by the unitarity constraints (7.2) on angular intensities using the expressions for $(R^j_k)^{JJ'}_{\lambda \lambda'}$ given in the Table I. and the parity relations (6.30) and (6.31) in the angular expansions of the intensities (6.32) and (6.33). We found that the combined constraints (7.9) for $k=y$ corresponding to $A_{yy}=-A_{uu}$ and $A_{uy}=0$ and the constraints $A_{xz}=0$ and $A_{xx}=A_{zz}$ are unconditional identities. In the Appendix we show that the constraint $A_{xx}=-A_{uu}$ holds true if and only if at least one of the following two constraints on the transversity amplitudes holds true   
\begin{equation}
\sum \limits_{J,\lambda \geq 0} \sum \limits_{K,\mu > 0} 
\eta_{\lambda}\eta_{\mu} \xi_{\lambda} \xi_{\mu} 
Im(U^J_{\lambda u}N^{K*}_{\mu d})ReY^J_\lambda(\Omega) ImY^K_\mu(\Omega) =0
\end{equation}
\[
 \sum \limits_{J',\lambda' \geq 0}\sum \limits_{K', \mu'> 0} 
\eta_{\lambda'}\eta_{\mu'}\xi_{\lambda'} \xi_{\mu'}
Im(U^{J'*}_{\lambda' d}N^{K'}_{\mu' u})ReY^{J'}_{\lambda'}(\Omega) ImY^{K'}_{\mu'}(\Omega)=0
\]
for all $\Omega=\theta,\phi$ and all $s,t,m$. The constraints $A_{ux}=0,A_{uz}=0,A_{xy}=0,A_{yz}=0$ are identities provided that both these constraints hold true. In (7.12) $\eta_{\lambda}=1,\xi_{\lambda}=1$ for $\lambda=0$ and $\eta_\lambda={1\over{\sqrt{2}}}, \xi_\lambda=2$ for $\lambda>0$. 

The constraints (7.12) imply constraints on the transversity amplitudes
\begin{equation}
Im(U^J_{\lambda \tau}N^{K*}_{\mu -\tau})=0
\end{equation}
that must hold true for all values of $J \lambda,K\mu,\tau$ at all values of kinematic variables $s,t,m$. To prove (7.13) we recall that at any dipion mass there is a finite number of contributing partial waves with $J \leq J_{max}(m)$ so that the sums (7.12) are truncated. Let $N$ be the number of the contribiting terms. We can select $N$ different values of $\Omega_i, i=1,N$ transforming (7.12) into a pair of $N$ linear homogeneous equations for the unknown $N$ terms given by the l.h.s. of (7.13). We can select the values of $\Omega_i$ such that the determinant of each system is non-zero which ensures that the unknown terms $(Im(U^J_{\lambda \tau}N^{K*}_{\mu -\tau}))_i,i=1,N$ must all vanish. 

\subsection{Unitary phases and their self-consistency}

For amplitudes with non-vanishing moduli the conditions (7.13) imply unitary relative phases between the amplitudes $U^J_{\lambda \tau}N^{K*}_{\mu -\tau}$
\begin{equation}
\Phi(U^J_{\lambda \tau})-\Phi(N^{K*}_{\mu -\tau})=0,\pm \pi, \pm 2\pi
\end{equation}
Keeping $N^{K*}_{\mu -\tau}$ or $U^J_{\lambda \tau}$ fixed, this condition implies unitary relative phases also between amplitudes  $U^J_{\lambda \tau} U^{J'*}_{\lambda' \tau}$ and $N^K_{\mu \tau}N^{K'*}_{\mu'\tau}$
\begin{equation}
\Phi(U^J_{\lambda \tau})-\Phi(U^{J'*}_{\lambda' \tau})=0,\pm \pi, \pm 2\pi
\end{equation}
\[
\Phi(N^K_{\mu \tau})-\Phi(N^{K'*}_{\mu' \tau})=0,\pm \pi, \pm 2\pi
\]
We shall refer to such $S$-matrix amplitudes as unitary amplitudes. It follows from (7.14) and the Table I. that all elements $R^0_z=R^2_z=0$ so that the intensities $I^0_z=I^2_z=0$.  

The relative phases of all $S$-matrix amplitudes must satisfy a phase condition
\begin{equation}
\Phi(A)-\Phi(B)=(\Phi(A)-\Phi(C))+(\Phi(C)-\Phi(B))
\end{equation}
for any triad of amplitudes $A,B,C$. As a result the relative unitary phases cannot be arbitrary but must form a self-consistent set satisfying (7.16). In addition, the relative phases must be in full accord with all measured interference terms. To build up such a self-consistent set we must start with the relative phases of $S$ and $P$-wave amplitues. For dipion masses $m \lesssim 980$ MeV the $S-P$ inteference terms require
\begin{subequations}
\begin{eqnarray}
\Phi(U^1_{0 \tau})-\Phi(U^0_{0 \tau}) & =\Phi(L_\tau)-\Phi(S_\tau)=0 \\
\Phi(U^1_{0 \tau})-\Phi(U^1_{1 \tau}) & =\Phi(L_\tau)-\Phi(U_\tau)=+\pi \\
\Phi(U^1_{1 \tau})-\Phi(U^0_{0 \tau}) & =\Phi(U_\tau)-\Phi(S_\tau)=-\pi
\end{eqnarray}
\end{subequations}
where we have introduced an alternate notation for the $S$ and $P$ wave ampltudes that will be used in the following Sections. These unitary phases define unitary phases in (7.14)
\begin{subequations}
\begin{eqnarray}
\Phi(U^0_{0 \tau})-\Phi(N^1_{1 -\tau}) & =\Phi(S_\tau)-\Phi(N_{-\tau})=0 \\
\Phi(U^1_{0 \tau})-\Phi(N^1_{1 -\tau}) & =\Phi(L_\tau)-\Phi(N_{-\tau})=0 \\
\Phi(U^1_{1 \tau})-\Phi(N^1_{1 -\tau}) & =\Phi(U_\tau)-\Phi(N_{-\tau})=-\pi
\end{eqnarray}
\end{subequations}
In the Table II. we present an example of a complete set of self-consistent unitary phases arising from these initial unitary phases. The relative phases involving $D$-wave and higher spin waves are hypothetical since the corresponding interference terms have not yet been directly measured. In the Table II. we define relative phase 
\begin{equation}
\omega=\Phi(S_d)-\Phi(S_u)=-(\Phi(N_d)-\Phi(N_u))
\end{equation}

\begin{table}
\caption{Self-consistent relative phases $\Phi(A)-\Phi(B)$ in bilinear products $AB^*$ of unitary amplitudes
$A=U^J_{0\tau},U^J_{\lambda\tau},N^K_{1\tau},N^K_{\mu\tau}$ and $B=U^{J'}_{0\tau'},U^{J'}_{\lambda'\tau'},N^{K'}_{1\tau'},N^{K'}_{\mu'\tau'}$
where $\lambda,\lambda'>0$ and $\mu,\mu'>1$. The superscripts $J,J'$ and $K,K'$ go with subscripts $0,\lambda,0,\lambda'$ and $1,\mu,1,\mu'$, respectively. The phase $\omega=\Phi(S_d)-\Phi(S_u)$.}
\begin{tabular}{|c|c|c|c|c|c|c|c|c|}
\toprule 
$AB^*$&$U^*_{0u}$&$U^*_{\lambda'u}$&$N^*_{1d}$&$N^*_{\mu'd}$&$U^*_{0d}$&
$U^*_{\lambda'd}$&$N^*_{1u}$&$N^*_{\mu'u}$\\
\colrule
$U_{0 u}$&0&$\pi$&0&$\pi$&$-\omega$&$-\omega+\pi$&$-\omega$&$-\omega+\pi$\\
\colrule
$U_{\lambda u}$&$-\pi$&0&$-\pi$&0&$-\omega-\pi$&$-\omega$&$-\omega-\pi$&
$-\omega$\\
\colrule
$N_{1d}$&0&$\pi$&0&$\pi$&$-\omega$&$-\omega+\pi$&$-\omega$&$-\omega+\pi$\\
\colrule
$N_{\mu d}$&$-\pi$&0&$-\pi$&0&$-\omega-\pi$&$-\omega$&$-\omega-\pi$&$-\omega$\\
\colrule
$U_{0d}$&$\omega$&$\omega+\pi$&$\omega$&$\omega+\pi$&0&$\pi$&0&$\pi$\\
\colrule
$U_{\lambda d}$&$\omega-\pi$&$\omega$&$\omega-\pi$&$\omega$&$-\pi$&0&$-\pi$&0\\
\colrule
$N_{1u}$&$\omega$&$\omega+\pi$&$\omega$&$\omega+\pi$&0&$\pi$&0&$\pi$\\
\colrule
$N_{\mu u}$&$\omega-\pi$&$\omega$&$\omega-\pi$&$\omega$&$-\pi$&0&$-\pi$&0\\
\botrule
\end{tabular}
\label{Table II.}
\end{table}

\section{Test of the unitary evolution law.}

\subsection{Unitary amplitude analysis of $S$- and $P$-wave subsystem at small t}

For dipion masses $m \lesssim 1080$ MeV and momentum transfers $|t| \lesssim 0.20$ (Gev/c$)^2$ the $S$- and $P$-waves dominate the $\pi^- p \to \pi^- \pi^+ n$ process. Using the Table I., the measured spin density matrix elements $R^0_u$ and $R^0_y$ organize into two groups of observables $a_{k,\tau}, k=1,6$ corresponding to target nucleon transversity $\tau=u$ and $\tau=d$. Their expressions in terms of transversity amplitudes in the notation introduced in the previous Section are given in the Table III.together with the expressions for the mesured elements $R^0_x$. Omitting the subscript $\tau$ for the sake of brevity, we find from the Table III.
\begin{subequations}
\begin{eqnarray}
|S|^2 = & a_1+a_2 -3|L|^2 \\
|U|^2 = & |L|^2-{1\over{2}}(a_2+a_3) \\
|N|^2 = & |L|^2-{1\over{2}}(a_2-a_3) \\
|L|^2 = & a_4a_5 /(a_6 \Gamma)
\end{eqnarray}
\end{subequations}
where
\begin{equation}
\Gamma={\cos(\Phi(L)-\Phi(S))\cos(\Phi(L)-\Phi(U))
\over{\cos(\Phi(U)-\Phi(S))}}=1
\end{equation}
\begin{table}
\caption{Measured spin observables for the $S$- and $P$-wave subsystem in terms of nucleon transversity amplitudes. The signs $+$ and $-$ correspond to $\tau=u$ and $\tau=d$, respectively. The superscripts $J=0,1$ are omitted. Real parts $Re R^0_k, k=u,y$ and imaginary parts $Im R^0_x$ are understood. The density matrix elements with subscripts $SP$ and $PP$ are defined by the equations (6.41)}
\begin{tabular}{|crl|}
\toprule
$a_{1,\tau}=$ & ${1\over{2}}((R^0_u)_{SP}\pm(R^0_y)_{SP})=$ & $|S_\tau|^2+
|L_\tau|^2 +|U_\tau|^2+|N_\tau|^2$ \\
$a_{2,\tau}=$ & $(R^0_u)_{PP}\pm(R^0_y)_{PP}=$ & $2|L_\tau|^2-|U_\tau|^2-
|N_\tau|^2$ \\
$a_{3,\tau}=$ & $(R^0_u)_{1-1}\pm(R^0_y)_{1-1}=$ & $|N_\tau|^2-|U_\tau|^2$ \\ 
$a_{4,\tau}=$ & ${1\over{2}}((R^0_u)_{0s}\pm(R^0_y)_{0s})=$ & $|L_\tau||S_\tau|
\cos(\Phi(L_\tau)-\Phi(S_\tau))$ \\
$a_{5,\tau}=$ & ${1\over{\sqrt{2}}}((R^0_u)_{01}\pm(R^0_y)_{01})=$ & $|L_\tau|
|U_\tau|\cos(\Phi(L_\tau)-\Phi(U_\tau))$ \\
$a_{6,\tau}=$ & ${1\over{\sqrt{2}}}((R^0_u)_{1s}\pm(R^0_y)_{1s})=$ & $|U_\tau||S_\tau|\cos(\Phi(U_\tau)-\Phi(S_\tau))$ \\
$r_1 =$       & $\sqrt{2}(R^0_x)_{s1}=$ & $-Re(S_u N^*_d)+Re(N_u S^*_d)$ \\
$r_2 =$       & $\sqrt{2}(R^0_x)_{01}=$ & $-Re(L_u N^*_d)+Re(N_u L^*_d)$ \\
$r_3 =$       & $(R^0_x)_{-11}=$        & $+Re(U_u N^*_d)-Re(N_u U^*_d)$ \\
\botrule
\end{tabular}
\label{Table III.}
\end{table}
For all $m$ below 1080 MeV $a_5<0$. For $m$ below 980 Mev $a_4>0$
and $a_6<0$. These signs yield the phases (7.17) and (7.18). For $m$ above 980 MeV $a_{4,d}$ and $a_{6,d}$ change signs prompting a change in relative phases
\begin{subequations}
\begin{eqnarray}
\Phi(L_d)-\Phi(S_d) & =+\pi \\
\Phi(L_d)-\Phi(U_d) & =+\pi \\
\Phi(U_d)-\Phi(S_d) & =0
\end{eqnarray}
\end{subequations}
and
\begin{subequations}
\begin{eqnarray}
\Phi(S_d)-\Phi(N_u) & =-\pi \\
\Phi(L_d)-\Phi(N_u) & =0 \\
\Phi(U_d)-\Phi(N_u) & =-\pi
\end{eqnarray}
\end{subequations}
The mixed sets of phases (7.17), (7.18) and (8.3), (8.4) for $\tau=u$ and $\tau=d$, respectively, still define a self-consistent set of relative phases.
In all cases $\Gamma=1$ so that the amplitudes $|L|^2$, and consequently all amplitudes in (8.1) as well, have a unique solution for both transversities. With $a_4=|L||S|$ and $a_5=\pm |L||U|$ we find from (8.1)
\begin{equation}
a_2=-a_1+3|L|^2 +{a_4^2\over{|L|^2}}
\end{equation}
\[
a_3=+a_1-|L|^2-{a_4^2+2a_5^2\over{|L|^2}}
\]
These variables thus are not independent and must be calculated during the analysis and tested for their being within the error volume of the data.

\begin{figure} {htp}
\includegraphics[width=12cm,height=10.5cm]{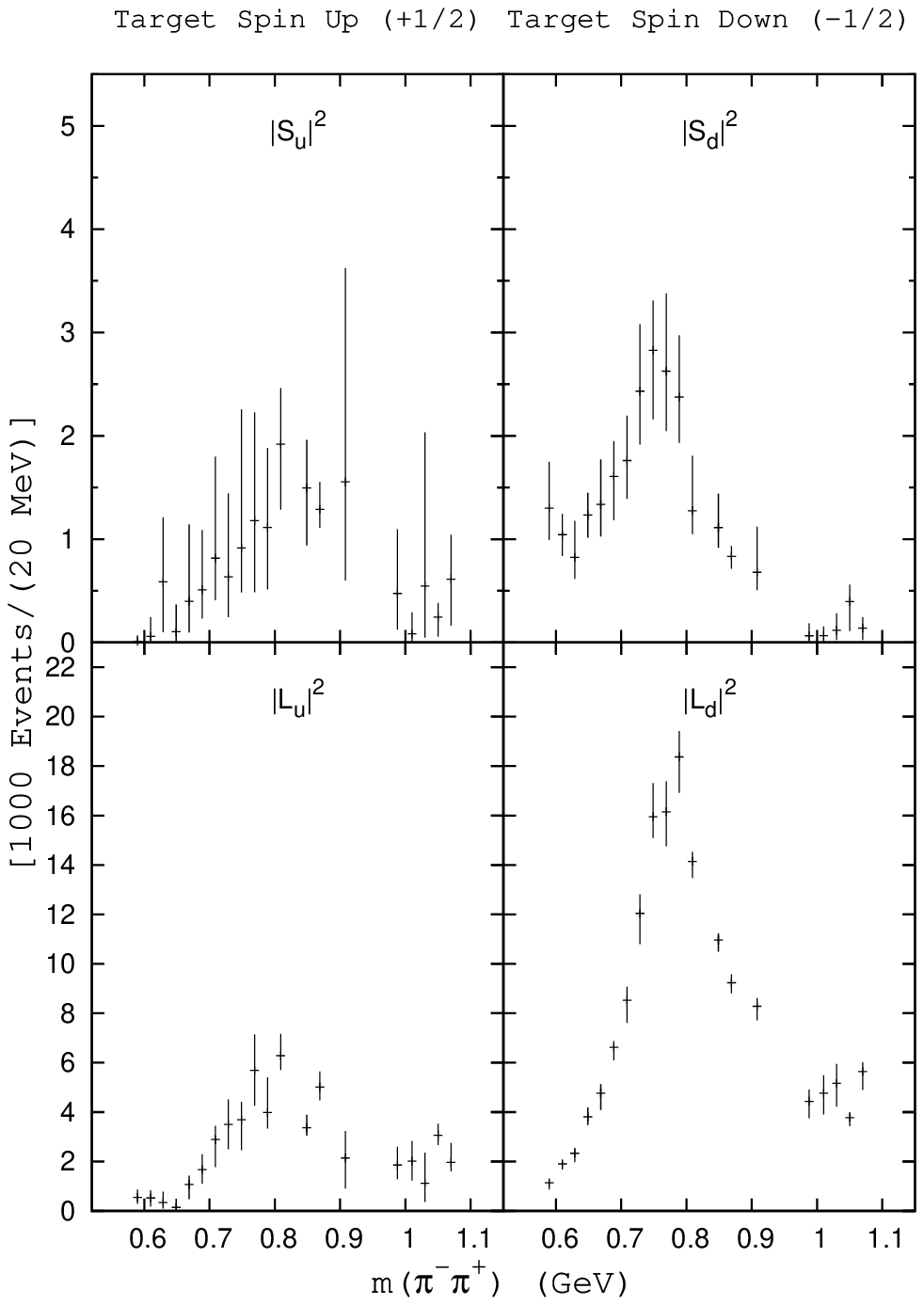}
\caption{$S$-wave and $P$-wave transversity amplitudes
$|S_\tau|^2$ and $|L_\tau|^2$ from unitary amplitude analysis.}
\label{Figure 1}
\end{figure}

\begin{figure} {hp}
\includegraphics[width=12cm,height=10.5cm]{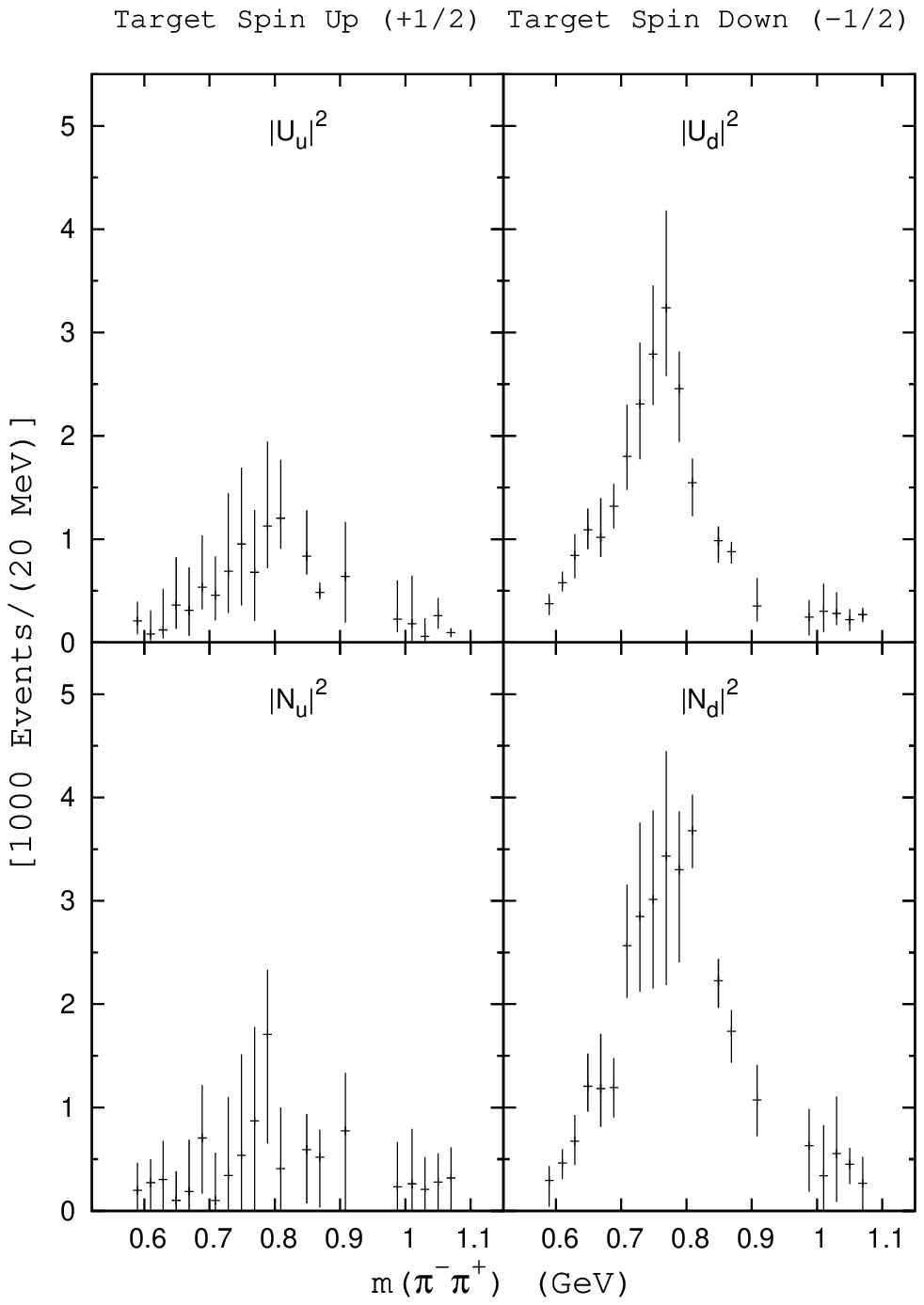}
\caption{$P$-wave transversity amplitudes
$|U_\tau|^2$ and  $|N_\tau|^2$ from unitary amplitude analysis.}
\label{Figure 2}
\end{figure}

\subsection{Unique solution for unitary amplitudes}

The amplitude analysis was carried out using a Monte Carlo method to search for physical solutions of amplitudes within the error volume of the data. The data were sampled using 10 million sampling data points. Initial analysis was unconstrained by fits to $R^0_x$ data and was followed by constrained analyses with fits to $R^0_x$ data constrained by 1, 3 and 5 standard deviations.

The unique solution for the moduli from the unconstrained analysis is shown in the Figures 1 and 2. The transversity "up" amplitudes are suppressed while the transversity "down" amplitudes dominate with a pronounced $\rho^0(770)$ peak. The data clearly require $\rho^0(770)$ in both $S$-wave amplitudes. There is a gap of no solution for masses 920-980 MeV in the $f_0(980)$ mass region suggesting the unitary phases are not compatible with $\rho^0(770)-f_0(980)$ mixing in the $P$-wave amplitude $|L_d|^2$ seen in the data in the previous analysis with non-unitary phases~\cite{svec07b}.

\begin{figure}
\includegraphics[width=12cm,height=10.5cm]{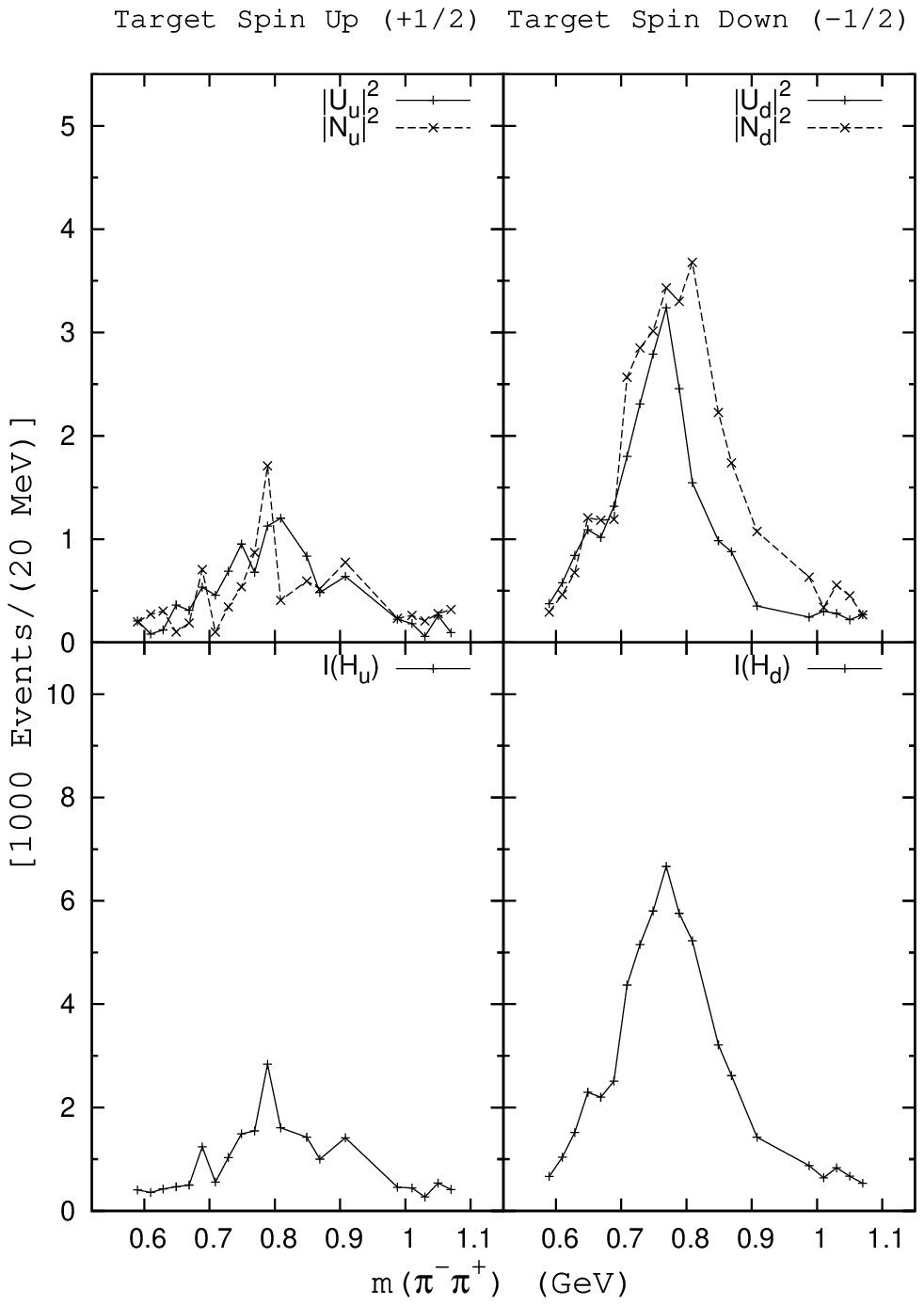}
\caption{The test of rotational/Lorentz symmetry of resonance production dynamics in $\pi^- p \to \pi^- \pi^+ n$ in amplitude analysis with unitary phases.}
\label{Figure 3}
\end{figure}

To test the rotational symmetry, and thus Lorentz symmetry, of the resonance production dynamics we need information on transversity amplitudes $H^\lambda_\tau$ with definite dipion helicity $\lambda=0,\pm 1$. The transverse amplitudes $U_\tau$ and $N_\tau$ are a mix of transverse amplitudes with helicities $\lambda = \pm 1$ and are thus not suitable to test the symmetry.
From (5.9) we find
\begin{equation}
H^{\pm 1}_\tau={1\over{\sqrt{2}}} (U_\tau \pm N_\tau)
\end{equation}
Their partial wave intensities can be calculated from the data on polarized target
\begin{equation}
I(H_\tau)=|H^{+1}_\tau|^2 + |H^{-1}_\tau|^2=|U_\tau|^2+|N_\tau|^2
\end{equation}
In Figure 3 we show the shape of the resonant peaks of $|U_\tau|^2$ and 
$|N_\tau|^2$. At half height the width of $|U_d|^2$ peak is $\sim 100$ MeV while that of $|N_d|^2$ is $\sim 180$ MeV. This difference arises from the interference terms of amplitudes $H^{\pm1}_d$
\begin{equation}
|U_d|^2={1\over{2}}(I(H_d)+P(H_d))
\end{equation}
\[
|N_d|^2={1\over{2}}(I(H_d)-P(H_d))
\]
where $P(H_d)=2Re(H^{+1}_d H^{-1*}_d)=|U_d|^2-|N_d|^2$. The figure shows the resonant shape of intensities $I(H_\tau)$. At half height the width of $I(H_d)$ is $\sim 150$ MeV - the proper width of $\rho^0(770)$ found also in the $\lambda=0$ amplitude $|L_d|^2$. These results support the rotational/Lorentz symmetry of the resonance production dynamics.

\subsection{Test of the unitary solution: Predictions for $R^0_x$} 

With unique moduli and unique phases (7.18) and (8.4), the unitary solution makes unique predictions for the measured elements $R^0_x$ given in the Table III.. For each sampling with physical solution we calculated the elements $R^0_u$, $R^0_y$ and $R^0_x$. For these predicted values we calclulated the corresonding value of $\chi^2$. From the distribution of the $\chi^2$ values we calculated their range of values and the average in each mass bin. From these average values of $\chi^2$ we calculated its bin averaged $\chi^2$ for each observable. In the case of $R^0_u$ and $R^0_y$ we further reduced the result to a more manageable average over the corresponding elements labeled $\chi^2 (<R^0_u>)$ and $\chi^2 (<R^0_y>)$, respectively. The goodness of the predictions is evaluated by (1) the number of empty mass bins with no physical solutions out of the total of 25 bins (2) the average number of "pass" events (i.e. physical solutions) per bin (3) the values of $\chi^2 (<R^0_u>)$ and $\chi^2 (<R^0_y>)$ (4) the bin average values of $\chi^2$  of the observables $R^0_x$ labeled $\chi^2((R^0_x)_{s1})$, $\chi^2((R^0_x)_{01})$ and  $\chi^2((R^0_x)_{-11})$. 

The Table IV. presents the results for 3 predictions. The first two predictions correspond to $\chi^2$ of $R^0_x$ data constrained in each mass bin to within 3 and 5 standard deviations for each sampling with a physical solution for the moduli. The third prediction is the result of unconstrained analysis with no such constraints on $\chi^2$ to $R^0_x$ data. All predictions for $R^0_u$ and $R^0_y$ have a good $\chi^2$. However, the first two predictions suffer from the large number of empty bins with no solution and a negligible number of constrained physical solutions in the remaining bins. The third prediction suffers from a very broad range of $\chi^2$ values for $R^0_x$ data in each bin and from very large bin-averaged values. This $\chi^2$ value is particularly large for the $S$-wave observable $(R^0_x)_{s1}$. We must conclude that the unitary solution is excluded by the $R^0_x$ data by at least 5 standard deviations. 

\begin{table}
\caption{Three sets of predictions of unitary amplitude analysis with $\chi^2$ for $R^0_x$ data constrained to within $3\sigma$, $5\sigma$ and unconstrained. Notation as defined in the text. Sampling size: 10 million data points.}
\begin{tabular} {|c|c|c|c|}
\toprule
prediction: & $3\sigma$ & $5\sigma$ & no constraint \\
\hline        
empty bins  & 11 of 25 & 9 of 25 & 5 of 25  \\  
\hline
pass/bin    & 1 & 1& 144,227 \\
\hline
$\chi^2(<R^0_u>)$ & 0.329 & 0.349 & 0.352 \\
\hline
$\chi^2(<R^0_y>)$ & 0.316 & 0.329 & 0.383 \\
\hline
$\chi^2((R^0_x)_{s1})$ & 0.891 & 1.278 & 6.494 \\
\hline
$\chi^2((R^0_x)_{01})$ & 0.718 & 0.737 & 3.762 \\
\hline
$\chi^2((R^0_x)_{-11})$ & 0.689 & 0.681 & 2.156 \\
\botrule
\end{tabular}
\label{Table IV.}
\end{table}

\subsection{The effect of dipion $D$-waves on unique unitary solution}

Measurements of $\pi N \to \pi \pi N$ do not measure the spin density matrix (sdm) elements $(R^0_k)^{JJ'}_{\lambda \lambda'}$ directly. Measurements on unpolarized targets measure moments $t^L_M$ and measurements on polarized target add moments $p^L_M$ and $r^L_M$~\cite{lutz78,svec12d,svec14a}. The moments $t^L_M,p^L_M,r^L_M$ are expressed in terms of $(R^0_u)^{JJ'}_{\lambda \lambda'},(R^0_y)^{JJ'}_{\lambda \lambda'},(R^0_x)^{JJ'}_{\lambda \lambda'}$, respectively. Only the moments $t^L_M$ and $p^L_M$ with $L \leq 2$ involve $S$- and $P$-wave sdm elements but they also include $S-D$ and $P-D$-wave interference sdm elements. $D$-wave sdm elements contribute only to moments $t^L_M$ and $p^L_M$ with $L=0,3,4$. There are $D-F$ interference terms in momemts with $L=3,4$. Explicit formulas for $t^L_M$ in terms of sdm elements are given in Ref.~\cite{grayer74}. The observables $a_{k,\tau}, k=1,6$ and $r_\ell,\ell=1,3$ in the Table III. thus include, in general, also $D$-wave contributions. 

Below the $K\bar{K}$ threshold for $m < 980$ MeV the $D$-wave moments with $L=3,4$ are very small compared to the $S$-and $P$-wave moments with $L=1,2$ with an exception of $t^3_1$ near $600-800$ MeV (Figure 14 of Ref.~\cite{grayer74}). Since this observation holds true in all other measurements of $\pi^- \pi^+$ and $\pi^+ \pi^-$ production, it is a common conclusion that for dipion masses $m<980$ MeV at low $t$ $S$-and $P$-wave amplitudes dominate and $D$-wave amplitudes can be neglected~\cite{becker79a,becker79b,chabaud83,rybicki85,svec92a,svec96,svec97a,svec07b,svec07c,svec12a,alekseev99}.

For $m>980$ MeV at low $t$ (from the $f_0(980)$ resonance) there is a sudden  increase in the momements with $L=3,4$ and the $D$-waves can no longer be neglected. Measurements on polarized targets in fact enable to determine the $D$-wave amplitudes (intensities) from $980-1600$ MeV in $20$ MeV bins. In the mass range $980<m<1080$ MeV the $D$-wave amplitudes are still relatively small compared to the $P$-wave which enables us to make a crude approximation of the $S$- and $P$-wave dominance even in this mass range. Figure 1 in Ref.~\cite{chabaud83} shows that the ratio of the $D$-wave intensities to the sum of the $S$-and $P$-wave intensities is ~$0.37$ at 990 MeV and ~$0.47$ at 1070 MeV. Since it is not possible to determine exact analytical solution for the $D$-wave amplitudes in terms of the measured polarization data certain approximations must be made in these analyses.

There are five $D$-wave transversity amplitudes: three unnatural exchange amplitudes $D^0_\tau,D^U_\tau,D^{2U}_\tau$ with helicities $\lambda=0,1,2$ and two natural exchange amplitudes $D^N_\tau,D^{2N}_\tau$ with helicities $\lambda=1,2$. The observables $a_{k,\tau}, k=1,6$ and $r_\ell,\ell=1,3$ can be written in the form
\begin{eqnarray}
a_{k,\tau} & = & c_{k,\tau}+d_{k,\tau}+e_{k,\tau}\\
\nonumber
r_\ell     & = & r_\ell(SP)+r_\ell(D)
\end{eqnarray}
where the $c_{k,\tau}$ are the $S$-and $P$-wave terms given in the Table III., $d_{k,\tau}$ are $D$-wave terms involving $D$-wave amplitudes with $\lambda \leq 1$
and $e_{k,\tau}$ are terms involving $\lambda=2$ amplitudes. Similarly, $r_\ell(SP)$ involve only the $S$- and $P$-wave terms given in the Table III. and 
$r_\ell(D)$ include D-waves. The expressions for $d_{k,\tau},e_{k,\tau},r_\ell(D)$
in terms of the transversity amplitudes are given in Ref.~\cite{svec12d,svec14a}.

\begin{figure} [htp]
\includegraphics[width=12cm,height=10.5cm]{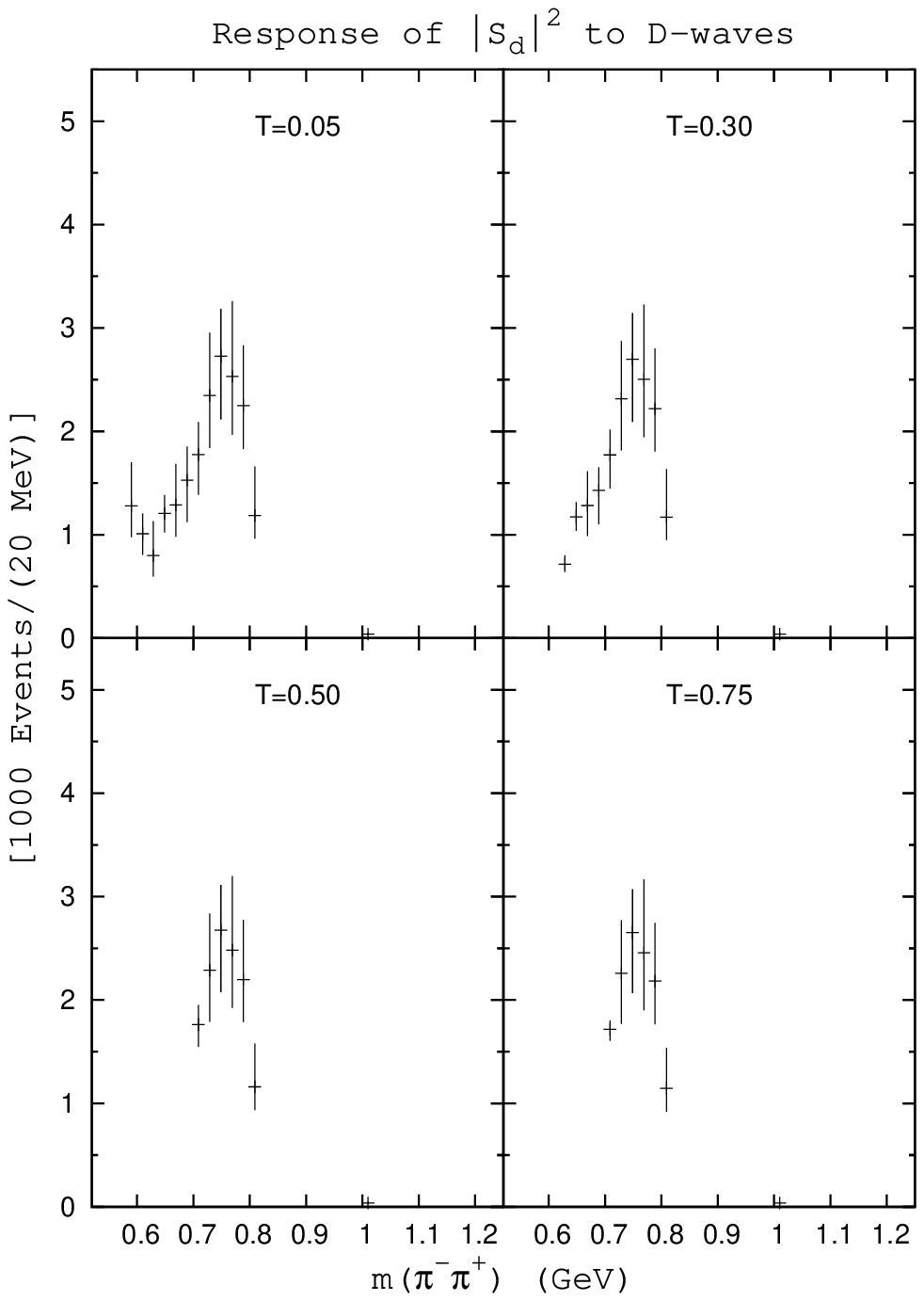}
\caption{Response of the amplitude $|S_d|^2$ to $D$-wave amplitudes in unitary analysis assuming $F=1.00$.}
\label{Figure 4}
\end{figure}

\begin{figure} [hp]
\includegraphics[width=12cm,height=10.5cm]{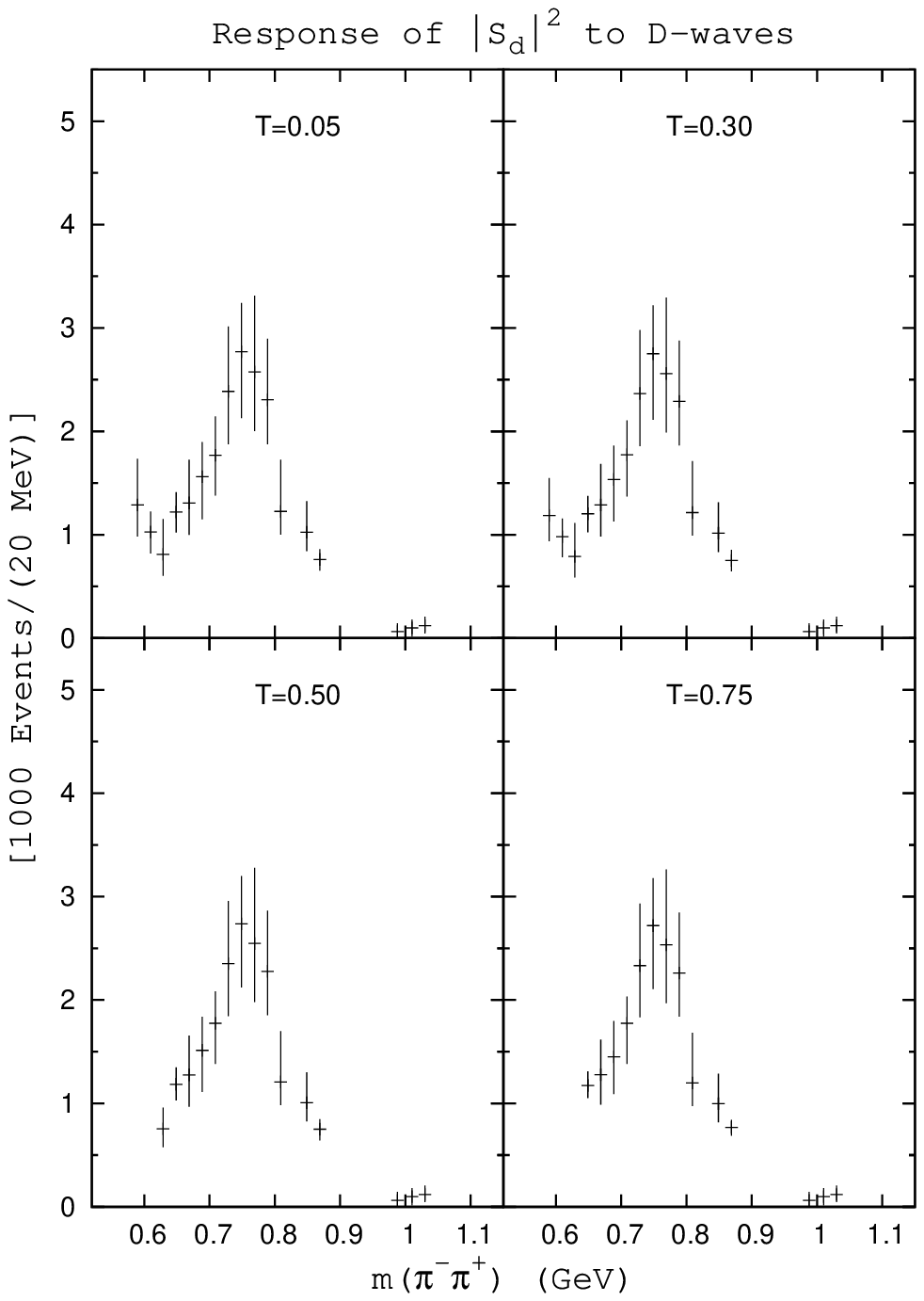}
\caption{Response of the amplitude $|S_d|^2$ to $D$-wave amplitudes in unitary analysis assuming $F=0.50$.}
\label{Figure 5}
\end{figure}

We wish to quantify the effect of the $D$-waves on the unitary amplitude analysis, both constrained and unconstrained. In our new analysis we shall neglect the smallest $D$ -wave amplitudes $D^{2U}$ and $D^{2N}$ and set $e_{k,\tau}=0$. For all relative phases we assume the consistent unitary phases given in the Table II. with the modifications (8.3) and (8.4). For the moduli $|D^0_\tau|, |D^U_\tau|, 
|D^N_\tau|$ we shall use a series of estimates obtained as follows to study the response of the unitary amplitudes.

For dipion masses $m>980$ MeV we know the $D$-wave intensities $I(A)=|A_u|^2+|A_d|^2$, $A=D^0,D^U,D^N$ from the amplitude analysis of the CERN measurement~\cite{chabaud83}. We have linearly extrapolated these intensities from their values $I_2(A)$ at $m_2=990$ MeV to value $I_1(A)=T I_2(A)$ at $m_1=590$ MeV where the fraction $T$ defines the slope parameter. The extrapolated intensities at mass $m$ are 
\begin{equation}
I(A,m)= T I_2(A)+\frac{(1-T)I_2(A)} {m_2-m_1}(m-m_1)
\end{equation}

Below 980 MeV there is a fairly constant ratio of the moduli $|A_u|^2:|A_d|^2 \approx 1:3$ for all $S$-and $P$-wave amplitudes. Using this ratio we have reconstructed the moduli of the $D$-wave amplitudes from the intensities
\begin{eqnarray}
|A_u(m)|^2=0.25I(A,m)F\\
\nonumber
|A_d(m)|^2=0.75I(A,m)F
\end{eqnarray}
where the factor $F$ accounts for the sudden decrease of $D$-wave moments below $980$ MeV. We varied the slope parameter $T$ in the range from 0.05 to 0.75 to estimate the $D$-wave amplitudes below 980 MeV. Above 980 MeV we have used the amplitudes (8.11) calculated from the measured intensities of the analysis~\cite{chabaud83}. The analysis was performed for $F=1.00$ and for $F=0.50$. We have used the conversion factor $0.109 \mu b/20 MeV = 1000 Events/20 MeV$ to convert the units of Ref.~\cite{chabaud83} to units of Ref.~\cite{grayer74} used in our analysis.

To accommodate the interference terms between the $D$-wave amplitudes and $S$- or $P$-wave amplitudes we used a form of a perturbation theory. Our code first calculated the unitary analysis assuming $c_{k,\tau}=a_{k,\tau}$ and, independently, the extrapolation of the $D$-wave aplitudes. The resulting $S$-and $P$-wave amplitudes were used to calculate the terms $d_{k,\tau}$. Then new "perturbed" $c_{k,\tau}=a_{k,\tau}-d_{k,\tau}$ were used to calculate new unitary $S$-and $P$-wave amplitudes to see their response to the presumed absence of the $D$-waves in the new parameters $c_{k,\tau}$. 

The results for the critical amplitude $|S_d|^2$ from the unconstrained analysis with $F=1.00$ at $T=0.05,0.30,0.50,0.75$ are shown in the the Figure 4. There are no physical solutions for $m > 800$ MeV and for $m<700$ MeV for $T=0.50$ and $T=0.75$ while there is only a little change in the $\rho^0(770)$ mass region at all $T$. These results suggests that the $D$-wave contribution is overestimeted in this unitary amplitude analysis which motivates us to consider the case $F=0.50$.  

The results for the critical amplitude $|S_d|^2$ from the unconstrained analysis with $F=0.50$ at $T=0.05,0.30,0.50,0.75$ are shown in the the Figure 5. Apart from an increase of empty bins from 5 to 8-11 at masses away from the $\rho^0(770)$ mass, this amplitude shows only a very weak response to $D$-waves at all $T$. In particular, there is no change in the presence of the $\rho^0(770)$ resonance in the $S$-wave. 

The Table V. quantifies the responses of the unitary analysis with $F=0.50$ constrained by $5\sigma$ fits to the data on sdm elements $R^0_x$ and of the analysis unconstrained by such fits. In the unconstrained analysis there is a modest improvemet in $\chi^2$ for $(R^0_x)_{s1}$ but which remains still too high. The constrained analysis has a higher number of empty bins at all $T$ than the constrained analysis without $D$-waves and a similarly negligible number of physical solutions per bin. On this basis we conclude that the unitary solution is excluded at the $5\sigma$ level with or without the $D$-wave contributions in the input data. 

\begin{table}
\caption{Four sets of predictions of unitary amplitude analysis including dipion $D$-waves $D^0,D^U,D^N$ assuming $F=0.50$ with $\chi^2$ for $R^0_x$ data constrained to within $5\sigma$ and unconstrained for parameter $T=0.05$ and $T=0.75$. Notation as defined in the text. Sampling size: 10 million data points.}
\begin{tabular} {|c|c|c|c|c|}
\toprule
prediction: & $5\sigma$ & $5\sigma$ & no constraint & no constraint \\
\hline
$T$         &    0.05  &   0.75   &   0.05  &   0.75  \\
\hline        
empty bins  & 12 of 25 & 14 of 25 & 8 of 25 & 11 of 25 \\  
\hline
pass/bin    &   1      &   106    & 121,190 &  80,498 \\
\hline
$\chi^2(<R^0_u>)$ & 0.379 & 0.380 & 0.351 & 0.370 \\
\hline
$\chi^2(<R^0_y>)$ & 0.360 & 0.365 & 0.383 & 0.394 \\
\hline
$\chi^2((R^0_x)_{s1})$  & 1.776 & 1.379 & 5.112 & 4.501 \\
\hline
$\chi^2((R^0_x)_{01})$  & 0.966 & 1.024 & 3.948 & 3.643 \\
\hline
$\chi^2((R^0_x)_{-11})$ & 0.830 & 0.641 & 2.452 & 2.601 \\
\botrule
\end{tabular}
\label{Table V.}
\end{table}

%\section{From non-unitary evolution to the quantum environment: the evidence}

\section{Evidence for the quantum environment and its pure dephasing interaction with particle scattering processes.}

\subsection{The hypothesis of the quantum environment}

In the Introduction we have put forward a hypothesis that the physical Universe includes a quantum environment which interacts with some particle scattering and decay processes. To be consistent with the conservation laws of the Standard Model this new interaction must be a pure dephasing interaction of the produced $S$-matrix final state $\rho_f(S)$ with quantum states $\rho(E)$ of the environment. It manifests itself by modifying (dephasing) the phases of the $S$-matrix amplitudes in a non-unitary evolution of the produced final state $\rho_f(S)$ to the observed final state $\rho_f(O)$ given by the Kraus representation. 

The most general form of Kraus representation reads 
\begin{equation}
\rho_f(O) = \sum \limits_{k=1}^M A_k \rho_f(S) A_k^+
\end{equation}
In quantum theory the non-unitary evolution law (9.1) can be always derived from a unitary co-evolution of the initial quantum system $S_i$ with a quantum environment $E$ by tracing out the environment in the joint final state $\rho_f(S_f,E)$. The unitary evolution is given by
\begin{equation}
\rho_f(S_f,E)=U\rho(E)\otimes\rho(S_i)U^+
\end{equation}
where the quantum state(s) of the enviroment are described by the density matrix~\cite{svec13b,svec14a}
\begin{equation}
\rho(E)=\sum \limits_{m,n}^M p_{mn}(E)|e_m><e_n|
\end{equation}
Here $|e_m>,m=1,M$ are $M$ orthonormal eigenstates describing the interacting degrees of freedom of the environment. Their number is limited by the dimensions of the Hilbert spaces~\cite{nielsen00} 
\begin{equation}
M=\dim{H(E)} \leq \dim{H(S_i)}\dim{H(S_f)}
\end{equation}
Assuming the conservation of the quantum numbers of the states $|e_m>$ by the evolution operator 
\begin{equation}
<e_k|U|e_m>=\delta_{km}<e_k|U|e_k>=\delta_{km}V_k
\end{equation}
the trace $\rho(S_f)=Tr_E \rho_f(S_f,E)$ reads
\begin{equation}
\rho(S_f)=\sum \limits_{k=1}^M p_{kk}V_k\rho(S_i)V_k^+
\end{equation}
In particle scattering processes we use the notation $\rho(S_i)=\rho_i(S)$ and $\rho(S_f)=\rho_f(O)$. 

Given (9.1) we can always enlarge the Hilbert space $H(S_i)$ to $H(E)\otimes H(S_i)$ with $\dim{H(E)}=M$, define a unitary evolution of this system and recover (9.6) as a trace over the " quantum environment" $H(E)$. Given (9.6) we always recover (9.1) with the replacement
\begin{equation}
\sqrt{p_{kk}}V_k \equiv A_k
\end{equation}
The two forms of the non-unitary evolution are equivalent provided $M$ satisfies (9.4). The quantum environment can be either an ancillary non-physical (mathematical) quantum environment, or it can be a real physical quantum environment~\cite{nielsen00}. Any interaction of a physical environment with a quantum system is described by (9.6). 

\subsection{The evidence for a pure dephasing non-unitary evolution in $\pi N \to \pi \pi N$}

The existence of a physical quantum environment and its pure dephasing interaction with particle scattering is supported by the following chain of evidence for the non-unitary evolution in $\pi N \to \pi\pi N$ processes and its dephasing character. We assume the non-unitary evolution law (9.1) without a reference to quantum environment but the evidence applies equally well for the non-unitary evolution law (9.6) assuming interaction with the quantum environment.\\

(A) {\it The observed amplitudes are not $S$-matrix amplitudes.}

Amplitude analysis of $S$- and $P$-wave subsystem with unitary relative phases yields a unique solution from the data on the observables $R^0_u$ and $R^0_y$ which fails to fit the experimental data on the observables $R^0_x$ at $5 \sigma$ level. The moduli of the transversity amplitudes in this analysis are nearly identical to the moduli found in amplitude analyses without unitary phases. There is therefore only one reason for this failure: the unitary relative phases. This means that the partial wave amplitudes describing the CERN data cannot be the $S$-matrix amplitudes and must have non-unitary phases. \\

(B) {\it Non-unitary evolution is involved in $\pi N \to \pi \pi N$ processes.}

All previous amplitude analyses of $S$- and $P$-wave subsystem in $\pi^- p \to \pi^- \pi^+  n$ at 17.2 GeV/c  ~\cite{becker79a,becker79b,chabaud83,rybicki85,svec92a,svec96,svec97a,svec07b,svec07c,svec12a} and at 1.78 GeV/c~\cite{alekseev99} as well as in $\pi^+ n \to \pi^+ \pi^- p$ at 5.98 and 11.85 GeV/c~\cite{svec92a,svec96,svec97a} found non-unitary relative phases $\Phi_{LS}$, $\Phi_{LU}$ and $\Phi_{US}$. The analyses~\cite{becker79a,becker79b,chabaud83,rybicki85} and the extension~\cite{svec12c} of recent analyses Ref.~\cite{svec07b,svec12a}  included the data on $R^0_x$ in their fits. The analyses of the $S$,$P$,$D$ and $S$,$P$,$D$,$F$ subsystems yield non-unitary relative phases at higher dipion masses and momentum transfers~\cite{becker79b,chabaud83,rybicki85}. Recent amplitude analysis of $\pi^- p \to \pi^0 \pi^0 n$ at 18.3 GeV/c also found non-unitary phases~\cite{gunter01}. The contrast between the predicted unitary relative phases and the observed non-unitary phases presents an unambigous evidence for a dephasing non-unitary evolution involved in the $\pi N \to \pi\pi N$.\\

(C) {\it The non-unitary evolution evolves the produced final state $\rho_f(S)$ into observed state $\rho_f(O)$.} 

The evidence for this claim comes from two independent features of the observed  amplitudes.

(i) In Ref.~\cite{svec12d} we present a survey of $S$-wave moduli and intensities from all amplitude analyses of the five measurements of $\pi^- p \to \pi^- \pi^+ n$ and $\pi^+ n \to \pi^+ \pi^- p$ on polarized targets. All these analyses provide a remarkably consistent evidence for a rho-like state in the $S$-wave of these processes which is not seen in $\pi^- p \to \pi^0 \pi^0 n$ amplitude analysis of high statistics measurement at 18.3 GeV/c~\cite{gunter01}. In Ref.\cite{svec12a} we identify this rho-like state with $\rho^0(770)$ resonance indicating a $\rho^0(770)-f_0(980)$ spin mixing in the $S$- and $P$-wave subsystem. As is the case with $\rho^0(770)-\omega(782)$ isospin mixing, the $\rho^0(770)-f_0(980)$ spin mixing requires a spin mixing interaction. Since there is no spin mixing interaction in the Standard Model, this non-standard interaction must originate in the non-unitary evolution involved in the $\pi^- N \to \pi^- \pi^+ N$ processes.

(ii) The analyses~\cite{svec07b,svec12a} established that the mass and the width of $\rho^0(770)$ resonance Breit-Wigner peak observed in all $P$-wave amplitudes does not depend on its helicity $\lambda$ as required by the rotational/Lorentz symmetry of the $S$-matrix. These results are similar to the results of the unitary analysis shown in the Figure 3. 

These two findings imply that the resonance production and the spin mixing have separate dynamical origins. There is no reason to assume that the production process is described by anything other than the $S$-matrix dynamics. While the observed amplitudes reveal non-unitary spin mixing, they are also consistent with the Lorentz symmetric $S$-matrix dynamics and its unitary evolution law. This is possible if and only if the non-unitary evolution evolves the produced final state $\rho_f(S)$ into the observed final state $\rho_f(O)$ as described by the Kraus representation (9.1) or (9.6). The dephasing non-unitary evolution is then a final state evolution described by the amplitudes of the Kraus operators.\\

(D) {\it The non-unitary evolution is a pure dephasing evolution.}

We now show that the non-unitary evolution is a pure (non-dissipative) dephasing evolution. Suppose such is not the case and the evolution is dissipative. As the result of the exchange of the four-momentum of the final state particles with the environment there is no conservation of the total four-momentum. As a result the measured four-momenta of the two produced pions will no longer be able to generate the Breit-Wigner shape of produced resonances in the observed amplitudes. There is also a breakdown of the conservation of the total angular momentum due to the exchange of angular momentum with the environment leading to the breaking of rotational and Lorentz symmetry by the produced resonances. The observed mass and the width of the distorted resonance like $\rho^0(770)$ will depend on its helicity, contrary to the observations. We must conclude that the non-unitary evolution (9.1) or (9.6) is a pure dephasing evolution that leaves all four-momenta of the final state particles intact.

\subsection{Evidence for a physical quantum environment}

The hypothesis of the existence of the quantum environment is validated (A) by the necessity to explain the physical origin of the non-unitary evolution as a physical process and (B) by identifying the quantum environment as a component of dark matter in a plausible model.\\

(A) {\it The non-unitary evolution as a physical process}

A non-unitary evolution of a quantum system described by (9.1) by itself does not necessarily require an interaction of the system with a physical environment. The chief difference between the non-unitary evolution law (9.1) and (9.6) is that the Kraus operators $A_k$ in (9.1) describe only the non-unitary evolution of the quantum system. In this form Kraus operators lack any other physical meaning and there is no physical limit on their number $M$. As well in this case the non-unitary evolution has no explicit physical origin. In contrast the Kraus operators $V_k$ in (9.6) describe both the non-unitary evolution of the system and its interaction with the environment which gives them a clear physical meaning. In this case the non-unitary evolution is generated by the interaction of the quantum system with the quantum environment .

The observed non-$S$-matrix amplitudes are a result of the non-unitary evolution law given by (9.1) or (9.6) and carry information about a new physics beyond the Standard Model. The non-unitary phases and spin mixing are generated by the Kraus operators $A_k$ or $V_k$ and are specific signatures of this new physics. We expect that the new physics originates in a new physical process, not in abstract operators like $A_k$ that lack a clear physical meaning and whose number $M$ is not physically restricted. This is particularly desirable in the case of the $\rho^0(770)-f_0(980)$ spin mixing which implies spontaneous violation of rotational/Lorentz symmetry in the observed $S$-and $P$-wave amplitudes~\cite{svec13b}. We expect such important new physical effect to have a new physical origin. 

The experimental evidence presented in the preceding Section does not distinguish between the non-unitary evolution law (9.1) without a reference to quantum environment and the non-unitary evolution law (9.6) generated by the interaction of the final state $\rho_f(S)$ with the quantum environment. However when we impose the condition (9.4) on the dimension $M$ then the non-unitary evolution law (9.1) is entirely equivalent to the non-unitary law (9.6) and we are free to describe the non-unitary evolution as an interaction of the system with a quantum enviromnment. This enables us to identify the new physical process and its new physics with the new pure dephasing interaction with the environment. 

In the sequel paper~\cite{svec13b} we show that the consistency of the pure dephasing interaction with the symmetries and conservation laws of the Standard Model in $\pi N \to \pi \pi N$ processes requires that it be a dipion spin mixing interaction. Its effect is the mixing of $S$-matrix partial wave amplitudes with different spins to form new observable partial wave amplitudes. The theory predicts  $\rho^0(770)-f_0(980)$ spin mixing in the $S$-and $P$-wave amplitudes in $\pi^- p \to \pi^- \pi^+ n$ in excellent qualitative agreement with the experimental results~\cite{svec07b,svec12a}. Quantitative agreement with the CERN data is presented in the new analysis using spin mixing mechanism~\cite{svec14a}. The consistency of this new interaction with the Standard Model also supports the necessity for a physical process behind the non-unitary evolution and thus for a physical quantum environment.

Kraus operators must inform us about such new process. They can only do so if they are interpreted as matrix elemets $V_k=<e_k|U|e_k>$ describing a co-evolution of the state $\rho_f(S)$ with the physical quantum environment in terms of its interacting degrees of freedom. What also distinguishes (9.6) from (9.1) is the explicit presence of the information about the quantum environment in (9.6) in terms of the probabilities $p_{kk},k=1,M$. As a result the non-unitary evolution law must take the form (9.6) generated by a physical pure dephasing interaction of the produced state $\rho_f(S)$ with the quantum state $\rho(E)$ of the environment.\\

(B) {\it Physical nature of the quantum environment: Dark matter.}

The necessary and sufficient condition for the quantum environment to be a real physical environment is that its interacting degrees with the quantum system be physical interacting degrees of freedom, not ancillary ones. Then the pure dephasing interaction of the quantum system with the quantum environment will be also a real physical interaction. To validate the hypothesis of the existence of the quantum environment we need to identify the physical degrees of freedom of the environment which will also identify the process that generates the non-unitary evolution.

The consistency of the pure dephasing interaction with the particle Standard Model suggests that the quantum environment has a universal presence in the Universe which manifests itself in astrophysical observations. Astrophysical observations provide a convincing evidence for the existence of dark matter and dark energy which are omnipresent environments in the Universe. Dark matter is characterized by non-standard interactions with baryonic matter. The quantum environment is characterized by mixed quantum states $\rho(E)$ given by (9.3) where the eigenstates $|e_m>$ describe its interacting degrees of freedomt. The pure dephasing interaction is also a non-standard interaction between the produced states $\rho_f(S)$ and the quantum states $\rho(E)$. In this aspect there is an obvious similarity between the dark matter and the quantum environment.

It is our conjecture that the quantum states $\rho(E)$ are particles of a distinct component of cold dark matter and that the pure dephasing interactions are its interactions with baryonic matter. The eigenstates $|e_k>$ could represent some new physical interacting degrees characterising this component of dark matter and its non-standard interactions with baryonic matter. But particle physics already knows of physical eigenstates with non-standard interactions: neutrino mass eigenstates. In Ref.~\cite{svec14a} we find that the dimension $M=4$. This provides a specific physical motivation to identify the four eigenstates $|e_m>$ with the four neutrino mass eigenstates $|m_k>$ including the new presumed light mass eigenstate $|m_4>$. We refer to the mixed states $\rho(E)$ as dark neutrinos. In contrast the three active neutrinos $\nu_e,\nu_\mu,\nu_\tau$ of the Standard Model and the new light sterile neutrino $\nu_s$ are pure states. All neutrinos engage in dephasing interactions and form the quantum environment. The light sterile neutrinos background can be interpreted as a hot dark matter.

Hot dark neutrinos were created in dephasing interactions of all flavour neutrinos with a variety of scattering processes in the early Universe and then redshifted to form warm and then a late cold component of cold dark matter. A careful analysis of the formation of galactic and large scale structures in the Universe indicates that most dark matter should be cold or warm at the onset of the  galaxy formation when the temperature of the Universe was about 1 keV. At these temperatures dark neutrinos still formed a hot dark matter and thus can account for only a part of the dark matter. A possible candidate for warm dark matter is sterile neutrino with mass 7.1 keV produced via lepton-number driven MSW (Mikheev-Smirnov-Wolfenstein) resonant conversion of active neutrinos near or at the Big Bang Nucleosynthesis epoch~\cite{shi99,abazajian02}. These sterile neutrinos are predicted to have a two-photon X-ray radiative decay at 3.55 keV~\cite{abazajian01,abazajian14}. An emission line at 3.55-3.57 keV was recently detected in X-ray spectrum of galaxy clusters in two independent observations~\cite{bulbul14,boyarski14}. These results suggest a multicomponent neutrino structure of dark matter with dark neutrinos one such component.

Interpreting the quantum environment as a component of dark matter endows it with a physical and material identity. The non-unitary evolution is generated by the interaction of the produced final states $\rho_f(S)$ with the cold dark neutrino component of dark matter and the light sterile and active neutrinos backgrounds. We elaborate on this model of the quantum environment in Ref.~\cite{svec14a}.

\section{A physical interpretation of the unitary and non-unitary\\ relative phases.}

It follows from the unitary conditions (7.14) and (7.15) that all unnatural exchange amplitudes $U^J_{\lambda \tau}$ as well as all natural exchange amplitudes $N^J_{\lambda \tau}$ share the same absolute phase up to an integer multiple of $\pi$
\begin{eqnarray}
\Phi(U^J_{\lambda \tau}) & = & \Phi(S_\tau)+\pi n(U^J_{\lambda \tau})\\
\nonumber
\Phi(N^J_{\lambda \tau}) & = & \Phi(N_\tau)+\pi n(N^J_{\lambda \tau})
\end{eqnarray}
The unitary relative phases thus imply that if there is a resonance at mass $m_R$ in a partial wave with $J=J_R$, then all contributing partial waves will have the same resonant phase near $m_R$. Mathematically, this does not necessarily imply that any of the partial waves with $J \neq J_R$ must resonate and show a resonant peak or a dip since the non-resonant moduli can still have resonant phases. Although the conditions (7.13) appear to allow such resonance mixing, it is excluded by the the symmetries of the $S$-matrix. 

The $S$-matrix amplitudes have a general form
\begin{eqnarray}
U^J_{\lambda \tau} & = & \exp{\Phi(S_\tau)}
[(-1)^{n(U^J_{\lambda \tau})}|U^J_{\lambda \tau}|+i0]\\
\nonumber
N^J_{\lambda \tau} & = & \exp{\Phi(N_\tau)}
[(-1)^{n(N^J_{\lambda \tau})}|N^J_{\lambda \tau}|+i0]
\end{eqnarray}
The phases $\Phi(S_\tau)$ and $\Phi(N_\tau)$ are functions of energy $s$, momentum transfer $t$ and dipion mass $m$. Near the resonant mass $m_R$ the phases give rise to the usual form of the partial wave amplitudes $A^{J_R}_{\lambda \tau}$ with $J=J_R$ in terms of Breit-Wigner amplitudes $a_{BW}^{J_R}(m)$ and a complex background $B^{J_R}_{\lambda \tau}$
\begin{equation}
A^{J_R}_{\lambda \tau}=< \pi^-\pi^+, J_R \lambda|T|R,\lambda>a_{BW}^{J_R}(m)
<R,\lambda \tau_n|T|0\tau> + B^{J_R}_{\lambda \tau}
\end{equation}
where $R$ is the resonance and $T$ is the transition matrix which describes the production and subsequent decay of the resonance $R$ in the partial wave amplitude $A^{J_R}_{\lambda \tau}$. The production and decay processes respect the conservation laws of the Standard Model within the amplitude $A^{J_R}_{\lambda \tau}$.

The unitary relative phases of the $S$-matrix amplitudes and the non-unitary relative phases of the observed amplitudes may have a simple and interesting physical interpretation. Consider complex standing waves on a string of the length $L$ 
\begin{equation}
y_n(x,t)=A_n(x) \cos(\omega_n t)=a_n \exp(ik_nx)\cos(\omega_n t) 
\end{equation}
arising from the superposition of two complex waves
\begin{equation}
y_n(x,t)  =  a_n \exp i(k_nx-\omega_nt) + a_n \exp i(k_nx+\omega_nt)
\end{equation}
where the integer $n \geq 1$. With the wavelegth $\lambda_n=n/2L$ the wave number $k_n=2\pi/\lambda_n=\pi n/L$. With tension $F$ and linear mass density of the string $\mu$ the angular frequency $\omega_n=(n/2L)\sqrt{F/\mu}$. The amplitude $a_n$ may depend on $\omega_n$. The real part $Re y_n(x,t)$ and the imaginary part $Im y_n(x,t)$ correspond to standing waves on the string open and closed at both ends, respectively
\begin{eqnarray}
Re y_n(x,t) & = & a_n \cos(k_nx)\cos(\omega_n t)\\
\nonumber
Im y_n(x,t) & = & a_n \sin(k_nx)\cos(\omega_n t)
\end{eqnarray}
At the far end $x=L$ the phase of the standing wave is given by $\Phi_n=k_nL=n\pi$. At $x=L$ the relative phases of the standing waves are
\begin{equation}
\Phi_n-\Phi_m=(n-m) \pi =0,\pm \pi, \pm 2\pi,...
\end{equation}
and their wave functions $y_n(L,t)$ are
\begin{equation}
y_n(L,t) = \cos(\omega_n t)[ (-1)^n|A_n(L)|+i0 ]
\end{equation}
A vibrating string in a vacuum (an empty space) does not generate sound waves. In a medium the vibrating string generates sound waves of the frequency $\omega_n$ but the sound waves are no longer standing waves. As a result two sound waves with frequencies $\omega_n$ and $\omega_m$ will no longer have a  relative phase $(n-m) \pi$ as the wavelegth changes in the medium. The relative phases will change for the sound waves in the medium.

The unitary relative phases of the $S$-matrix partial wave amplitudes $U^J_{\lambda \tau}$ and $N^J_{\lambda \tau}$ have the same relative phases as the vibrating string at $x=L$. We can write (10.2) for $U^J_{\lambda \tau}$ in the form
\begin{eqnarray}
Re U^J_{\lambda \tau} & = & \cos{\Phi(S_\tau)}
[(-1)^{n(U^J_{\lambda \tau})}|U^J_{\lambda \tau}|+i0]\\
\nonumber
Im U^J_{\lambda \tau} & = & \sin{\Phi(S_\tau)}
[(-1)^{n(U^J_{\lambda \tau})}|U^J_{\lambda \tau}|+i0]
\end{eqnarray}
and similarly for $N^J_{\lambda \tau}$. The comparison of (10.9) and (10.8) suggests the real and imaginary parts of these partial wave amplitudes are both akin to complex standing waves on strings open and closed at both ends. This similarity suggests to consider the $S$-matrix partial wave amplitudes as representations of some kind of complex dynamical (vibrational) modes confined to a finite dynamical region of a transient compound state. The modes are associated with the produced partial waves $|p_cp_d,J\lambda,\tau>$. In empty space these partial waves propagate without a change leaving the $S$-matrix amplitudes intact. In the presence of a quantum environment the propagating partial waves $|p_cp_d,J\lambda,\tau>$  are modified by the environment. This results in the modification of the $S$-matrix partial wave amplitudes into the observed partial wave amplitudes with the ensuing non-unitary relative phases. The quantum environment is responsible not only for the generation of the non-unitary phases but also for the observed $\rho^0(770)-f_0(980)$ mixing~\cite{svec13b,svec12d,svec14a}.  

\section{Conclusions and Outlook.}

We have presented a spin formalism that allowed us to construct the full final state density matrix $\rho_f(S)$ in $\pi N \to \pi \pi N$ processes in $S$-matrix theory. We have shown that the unitary evolution law imposes specific constrains on the relative phases of the transversity amplitudes $U^{J}_{\lambda \tau}$ and $N^{J}_{\mu \tau}$. The contrast between these predicted phases and the observed phases in all amplitude analyses of the pion production process presents an apparent violation of the unitary evolution law. Previous attempts to test unitary evolution law were framed as the test of the quantum mechanics itself. The central idea of this work is that an apparent violation of the unitary evolution law does not signal the breakdown of quantum theory. Rather it is an unambigous evidence for a non-unitary evolution of the produced final state $\rho_f(S)$ to the observed final state $\rho_f(O)$ generated by a new interaction of the state $\rho_f(S)$ with a quantum environment. To render the production mechanism accessible to experimental observation and to be consistent with the Standard Model this new interaction must be a pure dephasing interaction. This new non-standard dynamics represents a new physics beyond the Standard Model. 

We identify the quantum environment with dark neutrinos which form a distinct component of dark matter. This interpretation endows the quantum environment with a physical and material identity which connects it to the physics of the dark sector of the Universe. A part of the quantum environment are also the active and light sterile neutrinos backgrounds. We elaborate on this model of the quantum environment in Ref.~\cite{svec14a}. 

The spin formalism developed in this work and its consequences for the unitary phases of partial wave amplitudes apply equally well to a number of other meson production processes such as $K N \to K \pi N$, $\pi N \to \pi K \Lambda$, $\overline{K} N \to \pi \pi \Lambda$ and others. These processes could be measured on polarized target and the measurements with $\Lambda$ would also allow measurements of recoil $\Lambda$ polarization by its weak decays. Modern polarized targets reach high values of polarization and enable to select an arbitrary direction of the polarization vector~\cite{leader01}. Such experiments would provide new and independent tests of the unitary evolution law and advance our understanding of the quantum environment and its pure dephasing interactions with particle scattering processes.

\appendix

%\newpage
\section{Proof of the unitary evolution constraints on parity conserving transversity amplitudes.}

The unitary evolution constraints (7.12) on parity conserving transversity amplitudes follow from the constraint $A_{xx}=-A_{uu}$ which we can write in 
the form
\begin{equation}
(I^1_x)^2+(I^3_x)^2+(I^2_u)^2-(I^0_u)^2=
-(I^1_u)^2-(I^3_u)^2-(I^2_x)^2+(I^0_x)^2
\end{equation}
Expansions of the intensities involve $Re(Y^J_\lambda Y^{J'*}_{\lambda'})$ and $Im(Y^J_\lambda Y^{J'*}_{\lambda'})$ on the l.h.s. and r.h.s. of (A1), respectively. Using the expressions for $R^j_k$ in the Table I. in the intensities (6.32) and (6.33), relabeling of some terms, parity relations (5.16), and a relation for spherical harmonics $Y^L_{-M} (\theta, \phi) = (-1)^M (Y^L_{M} (\theta, \phi))^{*}$, the intensities on l.h.s. of (A1) read
\begin{equation}
I^1_x=-2 \sum \limits_{J,\lambda \geq 0} \sum \limits_{J',\lambda' \geq 0}
\eta_{\lambda} \eta_{\lambda'} \Bigl ( \xi_{\lambda} \xi_{\lambda'}
  Re(U^J_{\lambda u}U^{J'*}_{\lambda' d}) ReY^J_\lambda ReY^{J'}_{\lambda'}
-4Re(N^J_{\lambda u}N^{J'*}_{\lambda' d}) ImY^J_\lambda ImY^{J'}_{\lambda'}
\Bigr )
\end{equation} 
\[
I^3_x=+2 \sum \limits_{J,\lambda \geq 0} \sum \limits_{J',\lambda' \geq 0}
\eta_{\lambda} \eta_{\lambda'} \Bigl ( \xi_{\lambda} \xi_{\lambda'}
  Im(U^J_{\lambda u}U^{J'*}_{\lambda' d}) ReY^J_\lambda ReY^{J'}_{\lambda'}
+4Im(N^J_{\lambda u}N^{J'*}_{\lambda' d}) ImY^J_\lambda ImY^{J'}_{\lambda'}
\Bigr ) 
\]
\[
I^2_u=\sum \limits_{J,\lambda \geq 0} \sum \limits_{J',\lambda' \geq 0}
\eta_{\lambda} \eta_{\lambda'} \Bigl ( \xi_{\lambda} \xi_{\lambda'}
(Re(U^J_{\lambda u}U^{J'*}_{\lambda' u})-Re(U^J_{\lambda d}U^{J'*}_{\lambda'd})) 
ReY^J_\lambda ReY^{J'}_{\lambda'}
\]
\[
-4(Im(N^J_{\lambda u}N^{J'*}_{\lambda' u})-Im(N^J_{\lambda d} N^{J'*}_{\lambda'd})) ImY^J_\lambda ImY^{J'}_{\lambda'}\Bigr ) 
\]
\[
I^0_u=\sum \limits_{J,\lambda \geq 0} \sum \limits_{J',\lambda' \geq 0}
\eta_{\lambda} \eta_{\lambda'} \Bigl ( \xi_{\lambda} \xi_{\lambda'}
(Re(U^J_{\lambda u}U^{J'*}_{\lambda' u})+Re(U^J_{\lambda d}U^{J'*}_{\lambda'd})) ReY^J_\lambda ReY^{J'}_{\lambda'}
\]
\[
+4(Im(N^J_{\lambda u}N^{J'*}_{\lambda' u})+Im(N^J_{\lambda d} N^{J'*}_{\lambda'd})) ImY^J_\lambda ImY^{J'}_{\lambda'}\Bigr ) 
\]
where $\eta_{\lambda}=1,\xi_{\lambda}=1$ for $\lambda=0$ and $\eta_\lambda={1\over{\sqrt{2}}}, \xi_\lambda=2$ for $\lambda>0$. Using these expressions the l.h.s. of (A1) reads
\begin{equation}
A=-16\sum \limits_{J,\lambda \geq 0} \sum \limits_{J',\lambda' \geq 0}
 \sum \limits_{K,\mu > 0} \sum \limits_{K', \mu' > 0}C_{\lambda \lambda', \mu \mu'} \Bigl (Re(U^J_{\lambda u}U^{J'*}_{\lambda' u}N^K_{\mu u}N^{K'*}_{\mu' u})
+Re(U^J_{\lambda d}U^{J'*}_{\lambda' d}N^K_{\mu d}N^{K'*}_{\mu' d}) 
\end{equation}
\[
+2Re(U^J_{\lambda u}N^{K*}_{\mu d}U^{J'*}_{\lambda' d}N^{K'}_{\mu' u}) \Bigr )
ReY^J_\lambda ReY^{J'}_{\lambda'} ImY^K_\mu ImY^{K'}_{\mu'}
\]
where $C_{\lambda \lambda', \mu \mu'}=\eta_{\lambda} \eta_{\lambda'} \eta_{\mu} \eta_{\mu'} \xi_{\lambda} \xi_{\lambda'} \xi_{\mu} \xi_{\mu'}$ and where we used some relabeling and identities
\begin{equation}
\sum \limits_{J,\lambda \geq 0} \sum \limits_{J',\lambda' \geq 0} 
\eta_{\lambda} \eta_{\lambda'} \xi_{\lambda} \xi_{\lambda'}
Im(U^J_{\lambda \tau}U^{J'*}_{\lambda' \tau})ReY^J_\lambda ReY^{J'}_{\lambda'}=0 
\end{equation}
\[
\sum \limits_{K,\mu > 0} \sum \limits_{K', \mu' >0}
\eta_{\mu} \eta_{\mu'} \xi_{\mu} \xi_{\mu'}
Im(N^K_{\mu \tau}N^{K'*}_{\mu' \tau}) ImY^K_\mu ImY^{K'}_{\mu'}=0
\]
After a similar procedure the r.h.s. of (A1) can be brought to the form
\begin{equation}
B = A+16\sum \limits_{J,\lambda \geq 0} \sum \limits_{J',\lambda' \geq 0}
\sum \limits_{K,\mu > 0} \sum \limits_{K', \mu' > 0}C_{\lambda \lambda', \mu \mu'} 4Im(U^J_{\lambda u}N^{K*}_{\mu d})Im(U^{J'*}_{\lambda' d}N^{K'}_{\mu' u})
ReY^J_\lambda ImY^K_\mu ReY^{J'}_{\lambda'} ImY^{K'}_{\mu'}
\end{equation}
Since $A=B$ the equation (A5) immediately implies the conditions (7.12).

%\newpage

\end{document}